\documentclass[graybox,dvipsnames,usenames]{svmult}

\usepackage{mathptmx}       
\usepackage{helvet}         
\usepackage{courier}        
\usepackage{type1cm}        
\usepackage{makeidx}         
\usepackage{graphicx}        
\usepackage{multicol}        
\usepackage[bottom]{footmisc}

\usepackage{amssymb}

\usepackage{tocloft} 
\usepackage[linktocpage=true]{hyperref} 
\usepackage{xcolor}
\hypersetup{
    colorlinks,
    linkcolor={red!50!black},
    citecolor={blue!50!black},
    urlcolor={blue!50!black}
}

\usepackage[utf8]{inputenc}
\usepackage[english]{babel}
\usepackage[numbers]{natbib}

\newcommand{\nucl}[3]{ \ensuremath{ \phantom{\ensuremath{^{#1}_{#2}}} \llap{\ensuremath{^{#1}}} \llap{\ensuremath{_{\rule{0pt}{.75em}#2}}} \mbox{#3} } }

\newenvironment{lyxlist}[1]
{\begin{list}{}
{\settowidth{\labelwidth}{#1}
 \setlength{\leftmargin}{\labelwidth}
 \addtolength{\leftmargin}{\labelsep}
 }}
{\end{list}}

\newcommand{\xte}{{\it RXTE}}
\newcommand{\chandra}{{\it Chandra}}

\makeindex             

\begin{document}

\title*{Thermonuclear X-ray bursts}
\author{Duncan K. Galloway and Laurens Keek}
\institute{Duncan K Galloway \at Monash Centre for Astrophysics, School of Physics \& Astronomy, Monash University, Clayton VIC 3800, Australia, \email{duncan.galloway@monash.edu}
\and Laurens Keek \at CRESST and X-ray Astrophysics Laboratory NASA/GSFC, Greenbelt, MD
20771 USA, \email{lkeek@umd.edu}\\Department of Astronomy, University of Maryland, College Park, MD
20742 USA}
\maketitle

\abstract{
Type-I X-ray bursts arise from unstable thermonuclear burning of accreted fuel on the surface of neutron stars.
In this chapter we review the fundamental physics of the burning processes, and summarise the observational, numerical, and nuclear experimental progress over the preceding decade.
We describe the current understanding of the conditions that lead to burst ignition, and the influence of the burst fuel on the observational characteristics.
We provide an overview of the processes which shape the
burst X-ray spectrum, including the observationally elusive discrete spectral features.
We report on the studies of  timing behaviour related to nuclear burning, including burst oscillations and mHz quasi-periodic oscillations. 
We describe the increasing role of nuclear experimental physics in the interpretation of astrophysical data and models.
We survey the simulation projects that have taken place to date, and chart the increasing dialogue between modellers, observers, and nuclear experimentalists.
Finally, we identify some open problems with prospects of a resolution within the timescale of the next such review.
}

\tableofcontents

\section{Overview}

Thermonuclear (type-I) bursts are triggered by unstable ignition of accreted fuel on the surface of neutron stars (e.g. \cite{hvh75}). The first such events were detected  in 1975 from the ultracompact binary 3A~1820$-$30 (in the globular cluster NGC~6624) with the {\it Small Astronomy Satellite 3} ({\it SAS-3}) and the {\it Astronomical Netherlands Satellite}\/ ({\it ANS}; \cite[]{clark76,grindlay76}). 
Since then, a growing population of bursters (currently numbered at 110\footnote{\url{http://burst.sci.monash.edu/sources}}) has been observed by almost every major X-ray satellite, sometimes for a substantial fraction of their mission duration.
These events remain a high observational priority, as a key diagnostic of the nature of newly-discovered transient X-ray binaries. In addition, thermonuclear bursts have been increasingly exploited to constrain the fundamental properties of the host neutron stars, and to probe the accretion environment of the binary system.

In this chapter we focus on the research areas that have been of highest priority to both observers and theorists in the previous decade. These areas include 
the fundamentals of ignition of different types of bursts;
the study of ``long'' bursts, both intermediate-duration (He) bursts and superbursts (the latter thought to burn carbon, rather than the H/He fuel for short bursts);
the study of burst oscillations; 
the physics of the formation of the burst spectra, and the focus on deriving neutron-star parameters from them;
the searches for, and theoretical predictions of, discrete features in the burst spectra;
the phenomenon of mHz QPOs, and the link with steady burning;
the role of nuclear experimental physics, and the growing dialogue between the different communities;
progress in simulating bursts;
and
the growing role for bursts in probing the environment around the neutron star.

This chapter serves as an update to a series of thorough previous reviews, beginning in 1993 with \cite{lew93} and the subsequent work developed from it \cite[]{lew95}; the 1998 review \cite{bil98a}, covering the current state of the theory and the unexpected results from the {\it European X-ray Observatory Satellite} (EXOSAT); the 2006 review (published initially in 2003), with its focus on new phenomena discovered in the preceding decade, including burst oscillations \cite{sb03}; and the 2012 review of burst oscillations and related phenomena \cite{watts12a}.

\subsection{Theory of Burst Ignition and Nuclear Burning Regimes}

Soon after their discovery, 
measurements of the 
ratio of burst to accretion energy (the so-called $\alpha$-parameter)
established that
X-ray bursts are produced by
thermonuclear burning of accreted material \cite{Woosley1976,Maraschi1977}. 
Here we introduce the mechanism of runaway burning and present the classical picture of the different
burning regimes.

\subsubsection{Fuel Accretion from a Binary Companion Star}
\label{theory:accretion}

The fuel for X-ray bursts originates from the outer layers of a binary
companion star. Typically, it is expected to have a composition similar to the Sun,
predominantly hydrogen and helium, as well as small amounts
of CNO and other metals. An exception are the ultra-compact X-ray
binaries (UCXBs; e.g., \cite{intZand2007}) where most of the accreted
material is helium, with at most a small mass fraction of hydrogen
($\lesssim10\%$; \cite{2003Cumming}). The material is transferred by Roche-lobe overflow
to the neutron star via an accretion disk at rates up to the Eddington
limit, of $\dot{M}_{\mathrm{Edd}}=3.0\times10^{-8}\,M_{\mathrm{\odot}}\mathrm{yr^{-1}}\,(1+X)^{-1}(R/10\,\mathrm{km})$,
with $X$ the hydrogen mass fraction and $R$ the neutron star radius
(ignoring general relativistic corrections). $\dot{M}$ can vary by orders of magnitude
on time scales of days to decades. In transient systems, accretion  occurs  episodically, and
is effectively switched off  (``quiescence'') for long periods of time in between. We will see that the properties
of X-ray bursts depend on $\dot{M}$, and therefore a wide range of
bursting behaviour may be observed from a single accreting neutron
star, as its $\dot{M}$ changes with time.

Material falling on the neutron star is expected to rapidly spread across
its surface, due to the lateral pressure gradients imposed by the strong ($\approx10^{14}\ {\rm cm\,s^{-2}}$) gravity. 
The presence of accretion-powered pulsations in some 16 low-mass binaries (of which 8 are also burst sources; 
\cite{pw12,sk17}) 
implies that the magnetic field strength is sometimes sufficient
to confine the accreted fuel. 
For most bursting
sources the magnetic field  is thought to be dynamically unimportant.

\subsubsection{Runaway Thermonuclear Burning in a Thin Shell}
\label{sec:ignition}

Accretion grows the fuel layer on top of the neutron star until the
depth for thermonuclear ignition is reached. It is convenient to express
the depth of the layer as the column depth at radius $r$: $y(r)=\int_{r}^{\infty}\rho(r^{\prime})\mathrm{d}r^{\prime}$,
with $\rho$ the mass density. The column depth $y$ measures the mass above a unit area
at radius $r$, making $y$ a mass coordinate measured from the outside
in. Hydrostatic equilibrium in the layer can be expressed as $\mathrm{d}P=-g\rho\mathrm{d}r$,
with $P$ the pressure and $g$ the gravitational acceleration. Because
the envelope is relatively thin, $g$ can be approximated as being
constant in the outer layers. Hydrostatic equilibrium is then 
simplified to $P=-gy$, which shows that $y$ is also a pressure coordinate.

Compression increases the density and temperature in the fuel layer
with depth. A few meters below the stellar surface (from $y\simeq10^{8}\,\mathrm{g\,cm^{-2}}$ for hydrogen/helium), the conditions
for thermonuclear fusion can be achieved, and the neutron star envelope
is heated by nuclear burning.
The thin-shell instability (aided by  mild electron degeneracy) prevents
the neutron star envelope from cooling by expansion, and allows the fuel layer
to continue to heat up (see \S\ref{sec:runaway} for further discussion).

\begin{figure}
\includegraphics{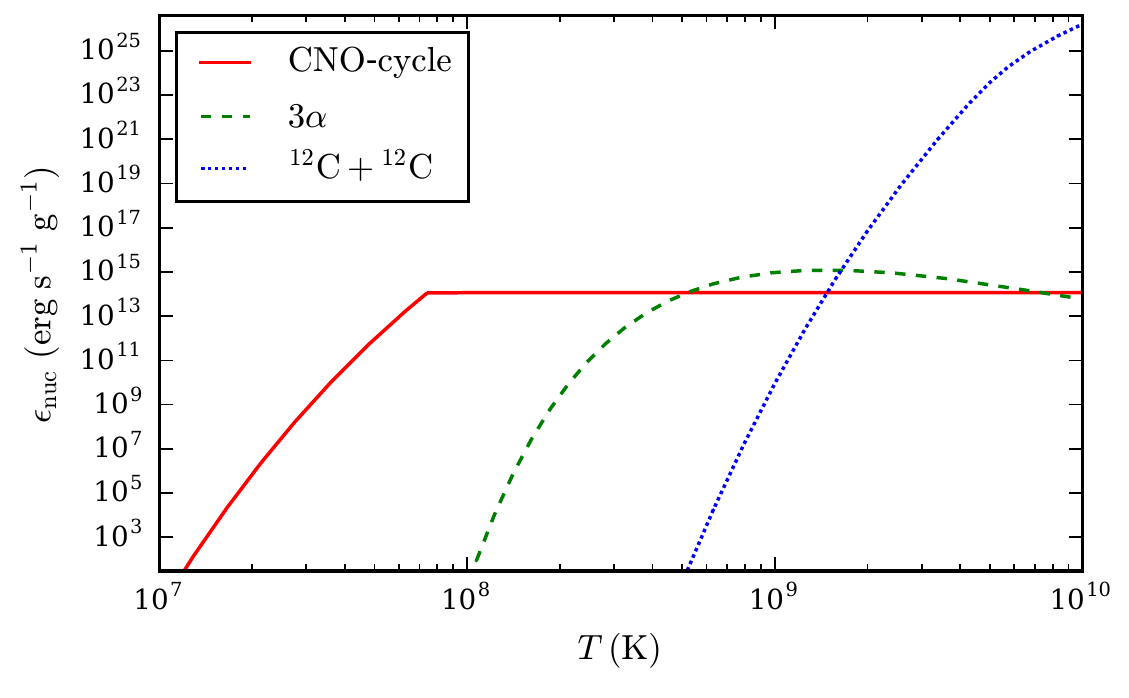}

\caption{\label{fig:enuc}The specific energy generation rate, $\epsilon_{\mathrm{nuc}}$, as
a function of temperature, $T$ for the three nuclear processes important for burst ignition. We calculate $\epsilon_{\mathrm{nuc}}$
at a density of $\rho=10^{5}\,\mathrm{g\,cm^{-3}}$ assuming a solar
composition ($X=0.73$, $Y=0.25$, $Z=0.02$) for the CNO cycle and
$3\alpha$, and $0.2$ of $^{12}\mathrm{C}$ for carbon fusion. For
CNO and $3\alpha$, $\epsilon_{\mathrm{nuc}}$ increases sharply with
$T$ at lower temperatures, which leads to a thermonuclear runaway.
At higher temperatures, the trend of $\epsilon_{\mathrm{nuc}}$ as
a function of $T$ flattens, and burning is stable.}
\end{figure}

For several of the important
nuclear processes, the thermonuclear burning rate increases sharply with
temperature (Fig.~\ref{fig:enuc}). Therefore, any heating from nuclear burning further increases
the burning rate. This leads to a \emph{thermonuclear runaway}, where
the burning rate increases quickly to the point where most fuel is locally
consumed within a second. 

\subsubsection{Ignition Conditions and the Stability of Burning}

The specific energy generation rate $\epsilon_{\mathrm{nuc}}$ 
for the three nuclear processes important for burst ignition
rises
steeply at lower $T$ and flattens at higher $T$ (Fig.~\ref{fig:enuc}). For helium burning
through the $3\alpha$ process, 
$\epsilon_{\mathrm{nuc}}$ 
depends weakly on $T$ above $T\gtrsim3\times10^{8}\,\mathrm{K}$, and stable
helium burning occurs instead of a thermonuclear runaway. Similarly,
hydrogen burning through the CNO cycle becomes stable at $T\gtrsim0.7\times10^{8}\,\mathrm{K}$.
These temperature thresholds are set by the depth of the Coulomb barrier which nuclei must overcome through quantum tunnelling. 
Because the H burning processes involve the conversion of protons into neutrons, they
include weak decays, which delay the process overall. 
The timescale for burning is set by the
$\beta$-decay of $^{14}\mathrm{O}$ and $^{15}\mathrm{O}$, with
half lives of $70.620\,\mathrm{s}$ and $122.24\,\mathrm{s}$, respectively.
At high temperatures H burning is known as the the ``hot CNO cycle'' or the ``$\beta$-limited CNO cycle'', where burning cannot run away and is,
therefore, stable \cite{ww81}.

\begin{figure}[ht]
\includegraphics[width=\textwidth]{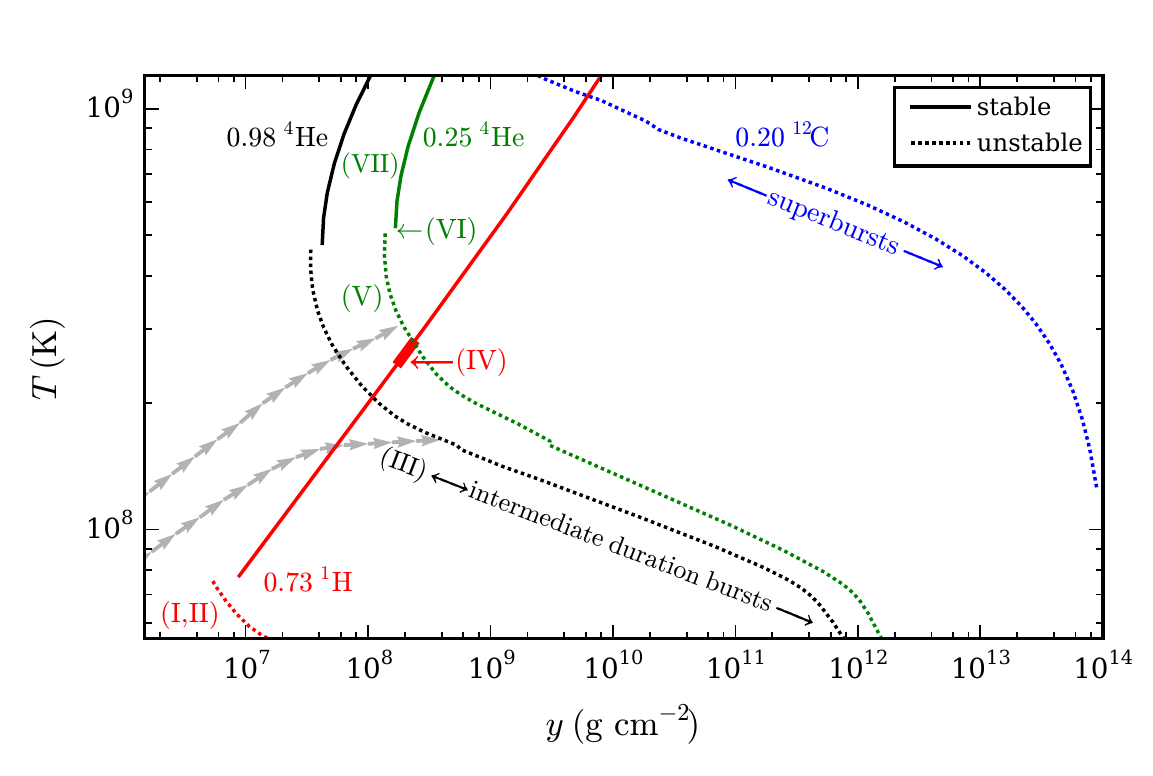}
\caption{Burning conditions as a function of column depth, $y$, and temperature, $T$, 
calculated with the one-zone model presented in \cite{Keek2016stable}.
Lines indicate the locations of stable burning and burst ignition (unstable burning) for three salient compositions: first, mixed hydrogen and helium at solar composition; second, a helium mass fraction of $0.98$ (for the case where all the hydrogen has burned prior to ignition); and third, typical ``superburst'' fuel containing $20\%$ carbon. Due to continual accretion, matter is compressed to higher $y$ and $T$ increases; lines of arrows (predicted by multi-zone models) exemplify two tracks that result in a mixed H/He burst (top) and a pure He burst (bottom). Labels indicate the burning regimes, with the roman numerals matching those in Table \ref{tab:burning_regimes}. The new stable H/He burning regime at sub-Eddington accretion rates (IV) is indicated by the thick red line.}
\label{fig:burning_regimes}
\end{figure}

Nuclear burning is only efficient
if it is stronger than radiative cooling, which is described by the
specific cooling rate $\epsilon_{\mathrm{cool}}=-\frac{acT^{4}}{3\kappa y}$,
with radiation constant $a$, speed of light $c$, and opacity $\kappa$
(e.g., \cite{Fujimoto1981,bil98a}). Stable burning in a steady-state
takes place at a depth where $\epsilon_{\mathrm{nuc}}=\epsilon_{\mathrm{cool}}$.
For thermonuclear runaways the ignition condition is $\frac{\mathrm{d}\epsilon_{\mathrm{nuc}}}{\mathrm{d}T}>\frac{\mathrm{d}\epsilon_{\mathrm{cool}}}{\mathrm{d}T}$,
which means that cooling cannot 
moderate small perturbations in the 
nuclear
burning rate. Both $\epsilon_{\mathrm{nuc}}$ and $\epsilon_{\mathrm{cool}}$
depend on $T$ and $y$, such that we can use these two conditions
to map out the ignition conditions for stable and unstable burning
(Fig.~\ref{fig:burning_regimes}; e.g., \cite{Fujimoto1981,fl87,bil98a,2003NarayanHeyl,Keek2016stable}).
For H and He, burning is unstable at lower $T$, and is stable at
higher values. Carbon burning is unstable over the entire considered
range of parameters\footnote{Multi-zone models find carbon burning to be stable in hot envelopes,
depending on $\dot{M}$ \cite{keek11}.}.

\subsubsection{Burning Regimes as a function of Accretion Rate}
\label{sec:regimes}

Which ignition conditions are reached on a neutron star, depends for
a large part on $\dot{M}$. The amount of heating due to compression
is proportional to $\dot{M}$. Therefore, $T$ is higher for larger
$\dot{M}$. Furthermore, steady-state burning requires that fuel is
burned at the same rate as at which it is accreted. For stable burning,
the temperature profile of the neutron star envelope adjusts to facilitate
this equilibrium.

\begin{table}
\begin{centering}
\caption{\label{tab:burning_regimes}Theoretical Nuclear Burning
Regimes$\,^{\mathrm{a}}$}
\begin{tabular}{ccl}
\hline 
$\dot{m}/\dot{m}_{\mathrm{Edd}}$ & & Burning Regime\tabularnewline
\hline 
 & (I) & Deep H flash (burns He)\tabularnewline
$\sim0.1\,\%\,^{\mathrm{b}}$ & & \tabularnewline
 & (II) & Shallow H flashes and deep He flash \tabularnewline
$0.\,4\%$  & & \tabularnewline
 & (III) & He flash (stable H burning) \tabularnewline
$8\,\%$ & & \tabularnewline
 & (IV) & Stable H/He burning \tabularnewline
$11\,\%$ & & \tabularnewline
 & (V) & Mixed H/He flash\tabularnewline
 & & \tabularnewline
$\sim100\,\%$$\,^{\mathrm{c}}$ & (VI) & Marginally stable burning of H/He\tabularnewline
 & & \tabularnewline
 & (VII) & Stable H/He burning\tabularnewline
\hline 
\end{tabular}
\par\end{centering}

$\,^{\mathrm{a}}$ For solar accretion composition and base flux  $Q_{\mathrm{b}}=0.1\,\mathrm{MeV\,u^{-1}}$ (see \S\ref{sec:base}; \cite{Keek2016stable}).

$\,^{\mathrm{b}}$ \cite{Peng2007}, including sedimentation.

$\,^{\mathrm{c}}$ \cite{Heger2005}. See also \cite{keek09,Zamfir2014,keek14b}.
\end{table}

The simple ignition conditions presented in Fig.~\ref{fig:burning_regimes} are determined
by considering hydrogen, helium, and carbon burning separately. Most
bursters accrete a mixture of hydrogen and helium, and the (typically steady) burning
of hydrogen influences the ignition of helium, via the heat that is contributed to the fuel layer. 
Steady H burning also affects the composition of the burst fuel at ignition, and hence the overall specific energy released by the burst, $Q_{\rm nuc}$. Where H is present in the burst fuel, the burning can proceed to heavier species via $(\alpha,p)$ reactions and proton captures (the rapid-proton, or {\sl rp}-process; e.g. \cite{schatz01}). $Q_{\rm nuc}$ can be inferred from measurements of the so-called $\alpha$ parameter, the ratio of the burst fluence to the peak flux (e.g. \cite{bcatalog}).

The variation of ignition and burning conditions for H and He 
leads to the prediction of a range of burning regimes 
as a function of $\dot{M}$
(Table~\ref{tab:burning_regimes};
\cite{Fujimoto1981,bil98a,Keek2016stable}), some of which have
been observed (\S\ref{burstobsstatus}):
\begin{lyxlist}{00.00.0000}
\item [{I}] At $T\lesssim7\times10^{8}\,\mathrm{K}$, hydrogen burns unstably.
The hydrogen-ignited flash quickly raises $T$. If the ignition depth,
$y_{\mathrm{ign}}$, is sufficiently large the ignition curve for
unstable helium burning is crossed, and helium burns along with hydrogen
in the burst.

\item [{II}] If $y_{\mathrm{ign}}$ is too shallow for runaway helium burning,
the hydrogen flash does not ignite helium. Instead, helium continues
to pile up until it reaches its ignition conditions at much larger
$y$. This regime, therefore, exhibits brief hydrogen flashes and
long helium bursts. (No observations matching case I or case II bursting have been identified). 

\item [{III}] At $T\gtrsim7\times10^{8}\,\mathrm{K}$, $\beta$CNO cycle
burning of hydrogen is stable. It heats the envelope and converts
hydrogen into helium. At $\dot{M}\lesssim0.08\,\dot{M}_{\mathrm{Edd}}$,
the buildup of the fuel layer is sufficiently slow for all hydrogen
to burn stably to helium. As most of the heating in the envelope comes
from the hydrogen burning, $T$ hardly increases with depth once hydrogen
is exhausted (see the lower series of arrows in Fig.~\ref{fig:burning_regimes}). Once
the helium ignition curve is reached, a pure helium burst ignites.

\item [{IV}] In a narrow range around $0.1\,\dot{M}_{\mathrm{Edd}}$,
heating by hydrogen burning creates the conditions for helium to burn
stably before reaching the helium ignition curve \cite{Keek2016stable}. This regime of stable hydrogen
and helium burning produces pure carbon ashes, which may in turn ignite
upon reaching the carbon ignition curve and power superbursts (see \S\ref{sub:Superbursts}).

\item [{V}] At higher $\dot{M}$, there is insufficient time for $\beta$CNO
burning to deplete hydrogen. Upon reaching helium ignition, hydrogen
is still present in the fuel layer, and it burns along with helium
in the burst. In this case burst ignition is more complicated because
of the interplay between hydrogen and helium burning. As $3\alpha$
burning of helium creates $^{12}\mathrm{C}$, the increase in the CNO
abundance promotes $\beta$CNO cycle burning of hydrogen. In turn, hydrogen
burning produces helium, which powers more $3\alpha$ burning. This mode
can be considered a combined runaway of hydrogen and helium burning.
Furthermore, once $T\gtrsim5\times10^{8}\,\mathrm{K}$, break-out
reactions from the $\beta$CNO cycle such as $^{15}\mathrm{O}\left(\alpha,\gamma\right)\mathrm{^{19}Ne}$
boost $\epsilon_{\mathrm{nuc}}$, and have an important effect on
the ignition conditions (\cite{Fisker2006,Davids2011,keek14b}).
See \cite{fis08} for a detailed description of the onset of
a mixed hydrogen/helium burst.

\item [{VI}] Near $\dot{M}_{\mathrm{Edd}}$ 
helium burning
is marginally stable due to competition
between the burning and cooling processes. It results in an oscillatory
burning mode (see \S\ref{sec:mHzQPOs}).

\item [{VII}] For $\dot{M}\gtrsim\dot{M}_{\mathrm{Edd}}$ the envelope
is sufficiently hot and fuel is accumulated fast enough for steady-state
(stable) helium burning. Hydrogen also burns stably.
\end{lyxlist}

If the accreted fuel does not contain hydrogen, helium flashes are expected to
occur at all mass accretion rates below the transition to stable helium
burning (regime VII). Carbon burning is further discussed in \S\ref{sub:Superbursts}.

\subsubsection{Base Heating, Rotational Mixing, and Gravitational Separation}
\label{sec:base}

The above theoretical picture successfully describes many of the observed burning regimes as a function of mass
accretion rate. The range of $\dot{M}$  where
each regime occurs observationally are, however, different from the predicted values. For example,
the transition from bursts to stable burning (around regime V) is observed
at $\sim 30\%\ \dot{M}_\mathrm{Edd}$ \cite{vppl88,corn03a,bcatalog}.
Furthermore, close to this transition
the observed bursts have relatively long or irregular recurrence times with indications of
stable burning in-between bursts (high $\alpha$-values, indicative of low $Q_{\rm nuc}$). This bursting regime 
does not match any of those in
Table~\ref{tab:burning_regimes} (see also \S\ref{burstobsstatus}). To improve the theoretical predictions, several
processes have been investigated in recent years.

First, a ``base heating'' term is included for
most models, which only describe the outer layers of the neutron star.
This additional flux, usually specified by the parameter $Q_\mathrm{b}$ (in
units of MeV per accreted nucleon), models the heat flowing from the underlying neutron star
crust into the envelope. The corresponding base luminosity,  $L_\mathrm{b}=\dot{M}Q_\mathrm{b}$, is thought to include contributions from pycnonuclear and electron-capture
reactions in the crust. Most simulations use a low value of $Q_\mathrm{b}=0.1-0.15\ \mathrm{MeV\ u^{-1}}$,
stemming from an early prediction \cite{Haensel1990}. However, more recent work indicates
that the generated heat may be larger, up to $Q_\mathrm{b}=2\ \mathrm{MeV\ u^{-1}}$ \cite{Haensel2003,Gupta2007}.
Additionally, a yet-unknown shallow heat source may increase $Q_\mathrm{b}$ \cite{bc09,Deibel2016}.

The degree of heating from below the fuel layer
$Q_\mathrm{b}$ may be reduced by the competing effect of neutrino cooling. Neutrino emissions from the core and crust strongly depend on temperature (e.g., \cite{cumming06}), and are therefore higher at larger $\dot{M}$: 
$Q_\mathrm{b}\simeq1.0\ \mathrm{MeV\ u^{-1}}$ at $0.01\ \dot{M}_\mathrm{Edd}$ and
$Q_\mathrm{b}\simeq0.1\ \mathrm{MeV\ u^{-1}}$ at $\ge 0.1\ \dot{M}_\mathrm{Edd}$ \cite{cumming06}.
Recently, a new Urca neutrino cooling process in the outer crust has been predicted, which
further complicates estimates of $Q_\mathrm{b}$ \cite{schatz14}. This parameter can affect the accretion rate range in which the burst regimes of \S\ref{sec:regimes} occur; for example, larger value of $Q_\mathrm{b}$ places
the transition between bursts and stable burning at a lower $\dot{M}$ \cite{keek09,Zamfir2014}.

Second, 
rotational mixing may be induced through 
the accretion of matter onto the neutron star, which includes the transfer
of angular momentum. This mixing is turbulent, and may involve a rotationally-induced magnetic field \cite{Piro2007,keek09}. Mixing transports the accreted fuel
more quickly down to the ignition depth, where it can burn in a stable manner. Therefore,
rotational mixing also reduces the $\dot{M}$ where the stability of nuclear burning is
predicted to change.

Third,
under the influence of the strong surface gravity, heavier isotopes
sink faster than lighter ones, leading to gravitational separation. At low mass accretion rates ($\dot{M}\lesssim0.01\ \dot{M}_\mathrm{Edd}$)
the burst recurrence times are sufficiently long for sedimentation of the fuel layer to
occur. When He and CNO separate from H, this influences the ignition conditions of H-ignited
bursts (regimes I and II in Table~\ref{tab:burning_regimes}) \cite{Peng2007}. Deeper in
the envelope, chemical separation may occur of carbon and heavier isotopes, which
influences the ignition conditions of superbursts \cite{Medin2011,Medin2014}.

Additional factors that may be important include
the ignition latitude (\S\ref{sec:latitude}),  
multi-zone and
multi-dimensional effects (\S\ref{sec:simulations}), and
the accuracy of nuclear physics data (\S\ref{sec:nuclear}).
These additional effects lead to much larger 
parameter-space for different ignition regimes than  the $\dot{M}$ 
dependence alone. Only a small part of this parameter space has been explored with numerical models
at this time.

\subsection{Status of burst observations}
\label{burstobsstatus}

Alongside the developments in theory have been commensurate improvements in our understanding of the phenomenology of bursts,  enabled in part by the assembly of large samples from long-duration X-ray missions.

\begin{figure}
\begin{center}
\includegraphics[width=0.8\textwidth]{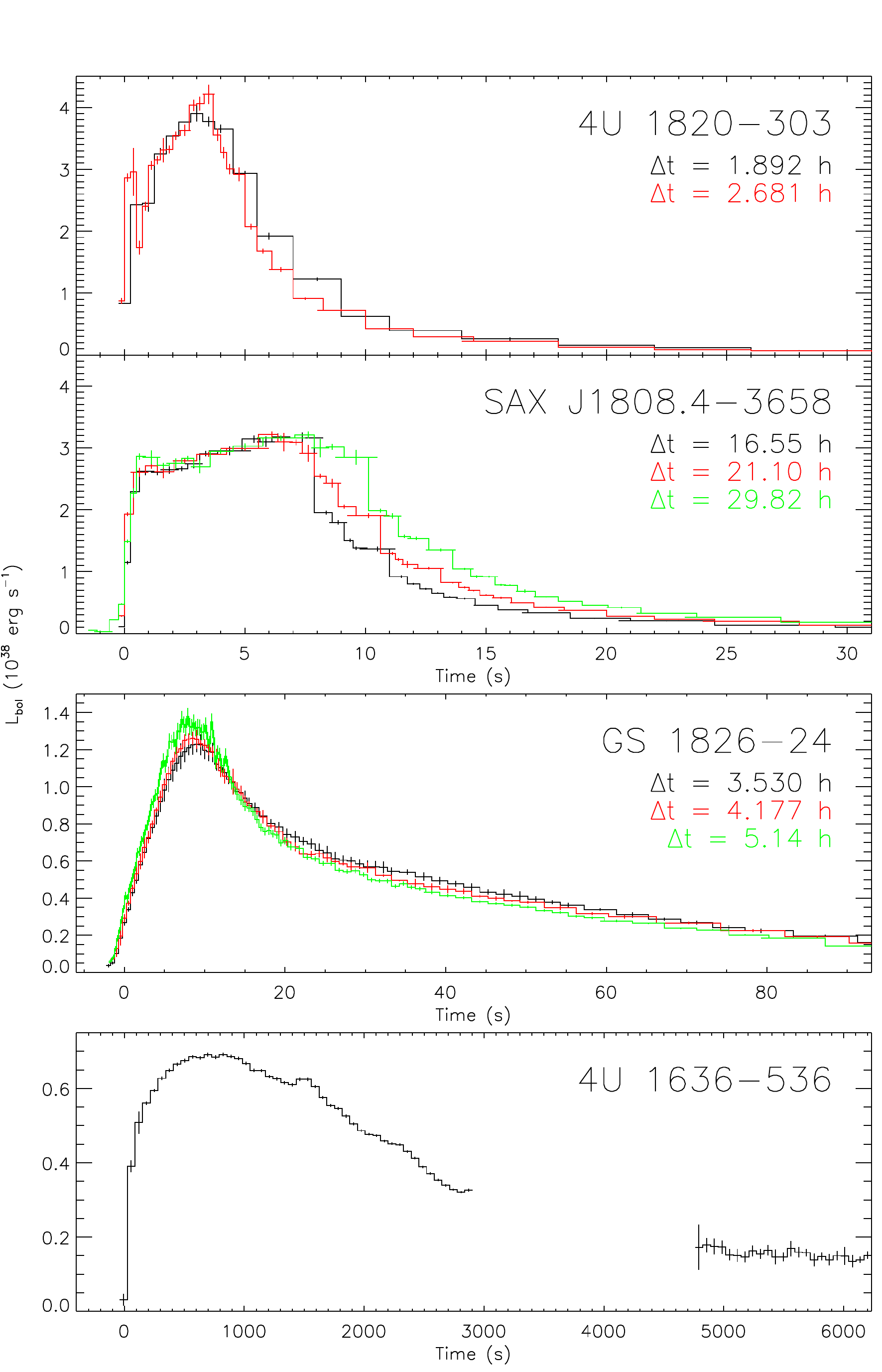}
\end{center}
\caption{Burst light curves observed by the {\it Rossi X-ray Timing Explorer}\/ (\xte) Proportional Counter Array (PCA) illustrating several different cases of  ignition, labelled with the source name and the recurrence time $\Delta t$, where known. The top panel shows likely pure He bursts ignited on the neutron star in the ultracompact system 3A~1820$-$30. The second panel shows He-rich bursts ignited after the accreted H has been exhausted at the base prior to ignition (case III in Table \ref{tab:burning_regimes}), in the accretion-powered millisecond pulsar SAX~J1808.4$-$3658. The third panel shows mixed H/He bursts from the ``Clocked burster'', GS~1826$-$24 (case V). The bottom panel shows an example of a superburst observed from 4U~1636$-$536.
Adapted from \cite{gal17a}. }
\label{fig:examples}
\end{figure}

\begin{figure}[ht]
\includegraphics[width=\textwidth]{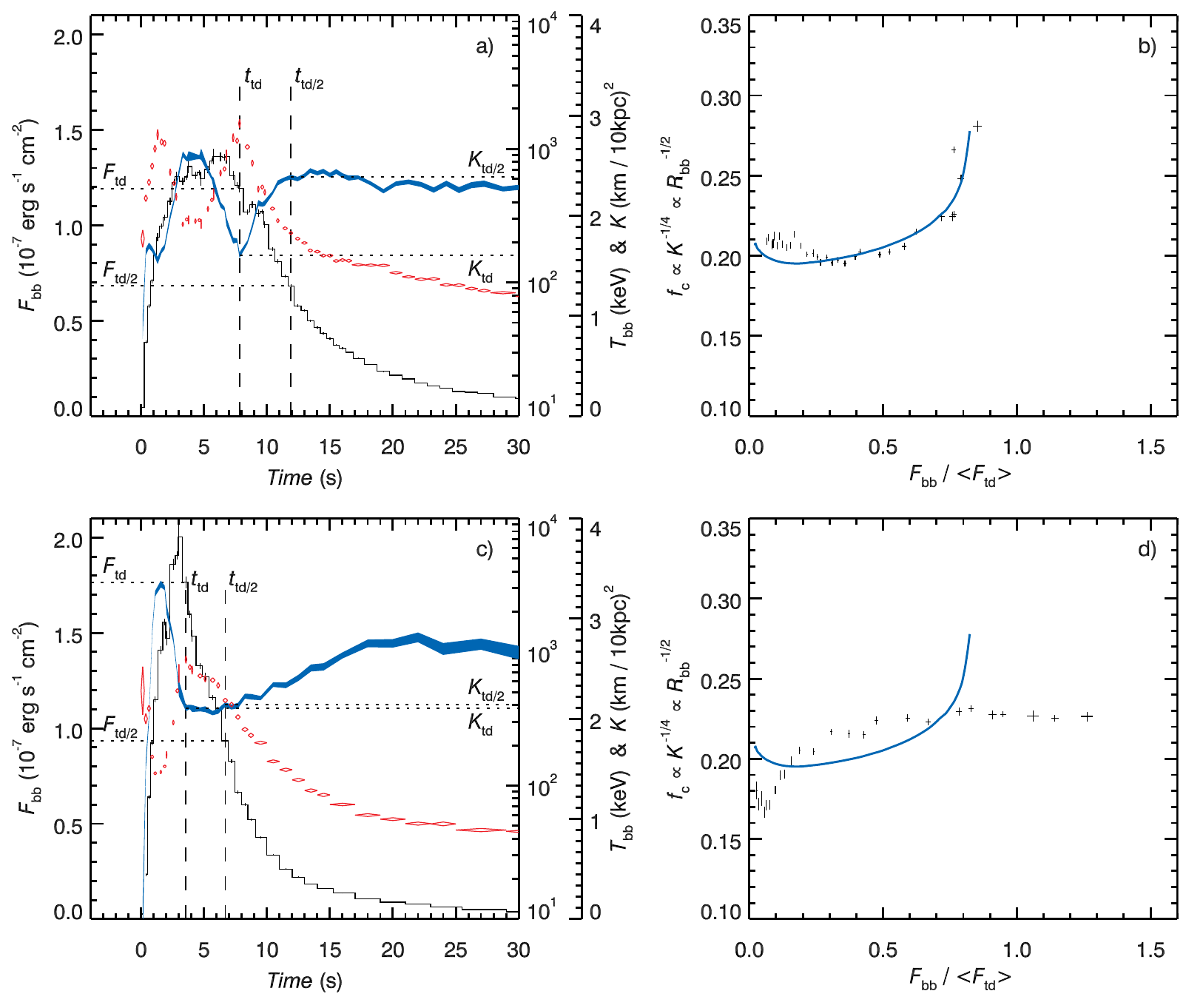}
\caption{Time-resolved spectroscopy of two X-ray bursts observed with \xte/PCA from the transient 4U~1608$–$52 illustrating the differences between hard state (top panels) and soft state (bottom panels) bursts. The left-hand panels (a and c) show the time-resolved spectroscopic parameters, with  
bolometric flux $F_\mathrm{bb}$ ({\it black line}, left-hand $y$-axis); 
blackbody normalisation $K_\mathrm{bb}$ ({\it blue ribbon}, inner right-hand $y$-axis); and 
blackbody temperature $T_\mathrm{bb}$ ({\it red diamonds}, outer right-hand $y$-axis). 
Note the  qualitative difference in the variation of $K_\mathrm{bb}$ during the two bursts (both events would be characterised as PRE).
The right-hand panels (b and d) show the relationship between the inverse square root of the blackbody radius (proportional to the colour-correction factor $f_\mathrm{c}$; see \S\ref{sec:continuum}) and the blackbody flux $F_\mathrm{bb}$ (scaled using the mean touchdown flux $F_\mathrm{td}$ for the source). The blue line (identical in both panels) is a model prediction for a pure hydrogen NS atmosphere with a surface gravity of $\log g = 14.3$, taken from 
\cite{suleimanov12a}.
The agreement between the atmosphere model and the data is much better for the hard state burst. 
Adapted from \cite{kajava14}. } 
\label{fig:burstspec}
\end{figure}

A clear distinction has been established between sources that accrete mixed H/He, and those which accrete pure (or almost pure) He, as from an ``ultracompact'' companion (e.g. \cite{bcatalog}). Sources accreting mixed H/He characteristically show regular, consistent bursts with long ($\approx5$~s) rise times and decays of a few minutes, understood to be powered by {\sl rp}-process burning (e.g. \cite{schatz01}; Fig.~\ref{fig:examples}). However, for most sources such bursts are only seen episodically, within the low/hard spectral state (see below; a notable exception is the so-called ``Clocked Burster'', GS~1826-24 \cite{clock99}).
At high accretion rates, mixed H/He accretors tend to show bursts with much shorter rise and decay times, often with burst oscillations (see \S\ref{oscillations}) and photospheric radius expansion (PRE; \cite{Tawara1984,Lewin1984}). This latter phenomenon is detectable by a temporary rise in the apparent (blackbody) radius (Fig. \ref{fig:burstspec}; see also \S\ref{sec:continuum}) around the peak of the burst, accompanied by a decrease in temperature sufficient to maintain a roughly constant luminosity. This behaviour is understood to arise when the burst luminosity reaches the (local) Eddington limit at the surface, resulting in a temporary expansion of the photosphere. When the energy input from the burning can no longer support the expanded photosphere, it falls back onto the star, resulting in a secondary increase in temperature (the ``touchdown point'') and followed by cooling at roughly constant radius.

Mixed H/He accretors are also the only sources that exhibit short recurrence time bursts (e.g. \cite{keek10}), weak events occurring just a few minutes after a brighter event. 
These events are significant because the interval since the previous burst is insufficient to reach the critical conditions for burst ignition  (see \S\ref{sec:swt}).

Burst sources where the accreted fuel is H-deficient, consistently show bursts characteristic of largely He-fuel, with short ($\lesssim1$~s) rise times and durations of 10--20~s (Fig.~\ref{fig:examples}; see also \S\ref{sec:long-duration-bursts}). One of the best-studied examples is 3A~1820$-$30, which consists of a neutron star in an 11-min orbit with an evolved companion (e.g. \cite{cumming06}). PRE bursts are more common, except at high accretion rates.

A secondary distinction has emerged between bursts that occur in the low (hard) persistent spectral state, and those in the high (soft) spectral state. These states 
are defined both by their spectral shape, and their periodic (and aperiodic) timing features (e.g. \cite{hvdk89}). The persistent spectral states are thought to indicate different characteristic accretion rate regimes; the low (hard) state, a truncated disk with the innermost accretion occurring through an optically thin, spherical flow; and the high (soft) state, with a disk extending to the NS surface and terminating in a boundary layer providing most of the persistent X-ray emission (e.g. \cite{done07}). Remarkably, these different accretion states also seem to give rise to markedly different burst behaviour. 

\paragraph{Low (hard) state bursts}

Thermonuclear bursts which occur when the persistent spectrum is hard (in colour-colour diagrams this state is referred to as the ``island'' state in {\it atoll} systems; e.g. \cite{hvdk89}) tend to exhibit long {\sl rp}-process tails, in systems that accrete mixed H/He. Radius-expansion bursts are relatively rare (e.g. \cite{muno01}), as are detections of burst oscillations (in systems where they have been seen). 
The burst rises tend to have a convex shape, suggesting ignition near the equator 
(see \S\ref{sec:latitude}).
The spectral evolution during these bursts shows a characteristic inflection which roughly matches the expectation of atmosphere models (e.g. \cite{suleimanov11a}; see also Fig.~\ref{fig:burstspec}). For this reason, these bursts have been adopted by some authors as the only reliable events from which the neutron star mass and radius can be inferred (e.g. \cite{kajava14}; see also \S\ref{sec:continuum}).

\paragraph{High (soft) state bursts}

When the persistent spectrum is soft (the ``banana'' state in colour-colour diagrams) the characteristics of bursts are remarkably different. Bursts are irregular, with longer recurrence times (on average), and exhibiting rapid ($\lesssim1$~s) rises and short ($\approx10$--20~s) durations characteristic of primarily He fuel, even in those systems which accrete mixed H/He. Radius-expansion is more frequently seen, as are burst oscillations. 
The burst rises tend to have a concave shape, suggesting the ignition location has moved away from the equator.
The blackbody normalisation is typically flat in the burst tail, which has motivated some observers to use the average as indicative of the projected radius \cite{guver12a}; but as this behaviour is at odds with the expected behaviour (arising from the flux-dependent distortion of the underlying blackbody component) this approach has raised some objections. 

\bigskip
We highlight this dichotomy here because the physical mechanism is not presently understood, but clearly has a wide impact on a range of investigations involving bursts. It is often assumed that the burst behaviour is independent of the details of the mechanism by which the accreted fuel arrives onto the neutron star (see \S\ref{theory:accretion}); the observations seem to suggest this cannot be true. 
That is, the observations suggest that the disk {\it geometry}\/ (inferred from the persistent spectral state) influences the burst physics, in addition to the accretion {\it rate}.
Alternatively, it is possible that the different accretion regimes give rise to different (local) conditions, perhaps allowing some ongoing level of steady He-burning. Such effects would certainly explain the weak bursts typically observed in the high (soft) state, as well as the decrease in burst rate at increasing accretion rates observed for many systems \cite{corn03a}.
As we will see, one of the growing areas of research over the past decade has been the interaction between the bursts and the accretion disk and environment of the neutron star (\S\ref{sec:interaction}).

    
\paragraph{The {\it Rossi X-ray Timing Explorer}}

Although many X-ray instruments have contributed to our present understanding of thermonuclear bursts,
a key instrument that has enabled much of the new results in burst phenomenology over the last decade
is the Proportional Counter Array (PCA; \cite{xte96}) on-board the {\it Rossi X-ray Timing Explorer}\/ ({\it RXTE}). This instrument, with an effective area of $\approx6500\ {\rm cm^2}$ and fast timing down to $1\mu s$ has allowed detailed studies of many prolific burst sources, with high signal-to-noise data permitting precise measurements of spectral and timing properties.
The assembly of the accumulated $\approx1100$ bursts over more than a decade of observations from more than 50 sources, led to a much greater capacity for broad-scale investigation into the burst phenomenology \cite{bcatalog}. Much of these investigations have focussed on the burst oscillation phenomenon, as described in \S\ref{oscillations}. The \xte\/ burst sample, as with earlier compilations (e.g. \cite{corn03a}) from other instruments, revealed a much more complex picture for unstable burning than predicted by theory.

An example is the study of photospheric radius-expansion bursts, which are thought to reach the local Eddington luminosity, and have been used as ``standard candles'' to determine the distance to bursting sources \cite{vp78}. Observations of bursts from globular cluster systems, which have an independently determined distance, allowed empirical measurement of the peak PRE burst luminosity, which is somewhat in excess of theoretical predictions for pure He fuel \cite{kuul03a}. Such studies also showed that in some systems there are substantial variations in the peak flux of PRE bursts from the same source. In 4U~1636$-$536, a small subset of PRE bursts reach a peak flux a factor of 1.7 lower than the majority of events, explained as a variation in the electron scattering dominated opacity between pure He and mixed H/He at solar composition.
This result, obtained for a sample of PRE bursts much larger than in any other source, further
suggests that typical PRE bursts reach the Eddington flux for 
material devoid of hydrogen,
even in systems accreting a mix of H and He \cite{gal06a}. In other systems, the typically 5--10\% variation in the peak fluxes may be attributed to interactions with the accretion disk \cite{gal03b}. Eddington-limited bursts remain the primary source of information on the distances to burst sources, but more work is required to reduce the effect of systematic contributions to the peak flux distribution.

Independently of the studies that accompanied the catalogue data itself, the provision of an index (i.e. burst times and observation IDs) into an otherwise undifferentiated collection of burst observations enabled other researchers to perform their own analyses on the burst sample, further increasing its impact.
The \xte\/  sample has been extended with further observations through to the end of the mission in 2012 January, and now totals more than 2000 bursts. These data will be incorporated into a new, and more comprehensive sample of bursts, the Multi-INstrument Burst ARchive (MINBAR\footnote{\url{http://burst.sci.monash.edu/minbar}}).
Together with burst observations of the {\it BeppoSAX}/WFCs \cite{corn03a}, and {\it INTEGRAL}/JEM-X, MINBAR contains over 7000 bursts.

\section{X-ray Burst Ignition}
\label{sec:runaway}

In \S\ref{sec:ignition} we explained burst ignition as due to runaway thermonuclear burning. 
Here we discuss the other fundamental processes that facilitate the ignition of different types of
bursts, including the so-called short recurrence time bursts, as well as the role of the
latitude of the ignition location.

\subsection{Thin-Shell Instability and Electron Degeneracy}
\label{sec:thinshell}

The typical response of a star to heating
is expansion, which reduces the pressure and thereby the temperature. For runaway nuclear
burning to occur, this cooling mechanism must be circumvented,
via a combination of the
thin-shell instability and electron degeneracy. Some past reviews highlight the latter \cite{lew93}, whereas others 
present the former \cite{sb03}. Here we review the two mechanisms.

X-ray bursts are an example of nuclear burning that happens in a thin shell \cite{Schwarzschild1965,Yoon2004a}.
If the radial extent of the burning layer is small with respect to
the stellar radius, expansion of the thin shell does not result in
a sufficient change in pressure to reduce the temperature. This allows
for the temperature to continue to increase, and is known as the \emph{thin-shell
instability}. For X-ray bursts and superbursts the ignition layer
has a depth of a few meters up to $\sim100$~m, which is very
thin compared to the $\approx10\,\mathrm{km}$ neutron star radius. Therefore,
the thin-shell instability allows for a growing temperature increase due to
burning, which leads to the ignition of bursts.

Electron degeneracy, identified
as the primary mechanism for the ignition of novae on white dwarfs
\cite{Giannone1967}, also plays a role in bursts on neutron stars. 
In degenerate material, the pressure is independent
of the temperature, such that a reduction in pressure does not counter
a temperature increase. On neutron stars, however, the electrons are
only mildly degenerate near $y\simeq10^{8}\,\mathrm{g\,cm^{-2}}$: the ratio of the electron
Fermi energy and thermal energy prior to a burst is $\eta \simeq 3$. A small amount
of heating at the burst onset lifts the degeneracy, but the runaway continues.

A simple numerical experiment elucidates the relative importance of the two
mechanisms. Employ a one-zone model to simulate the burning behaviour with and without
a pressure term for degenerate electrons\footnote{For example, the one-zone helium ignition model available at \url{https://github.com/andrewcumming/onezone}.}. When ignoring the pressure term from electron degeneracy, bursts are still
produced. Of course, electrons do contribute to the pressure, and ignoring that term
leads to a slightly larger ignition depth. Therefore, the primary mechanism for the
thermal instability of bursts is the thin-shell instability, while the electron degeneracy is important
for obtaining accurate ignition conditions.

Intermediate duration bursts and superbursts ignite deeper in the envelope (Table~\ref{tab:durations}), where
the electron degeneracy is larger, up to $\eta \simeq 10^2$. Nevertheless, their fuel
layer is a thin shell, which implies that long duration bursts would also occur without
the degeneracy effect.

\subsection{Reignition after a Short Recurrence Time}
\label{sec:swt}

\begin{figure}
\includegraphics[width=\textwidth]{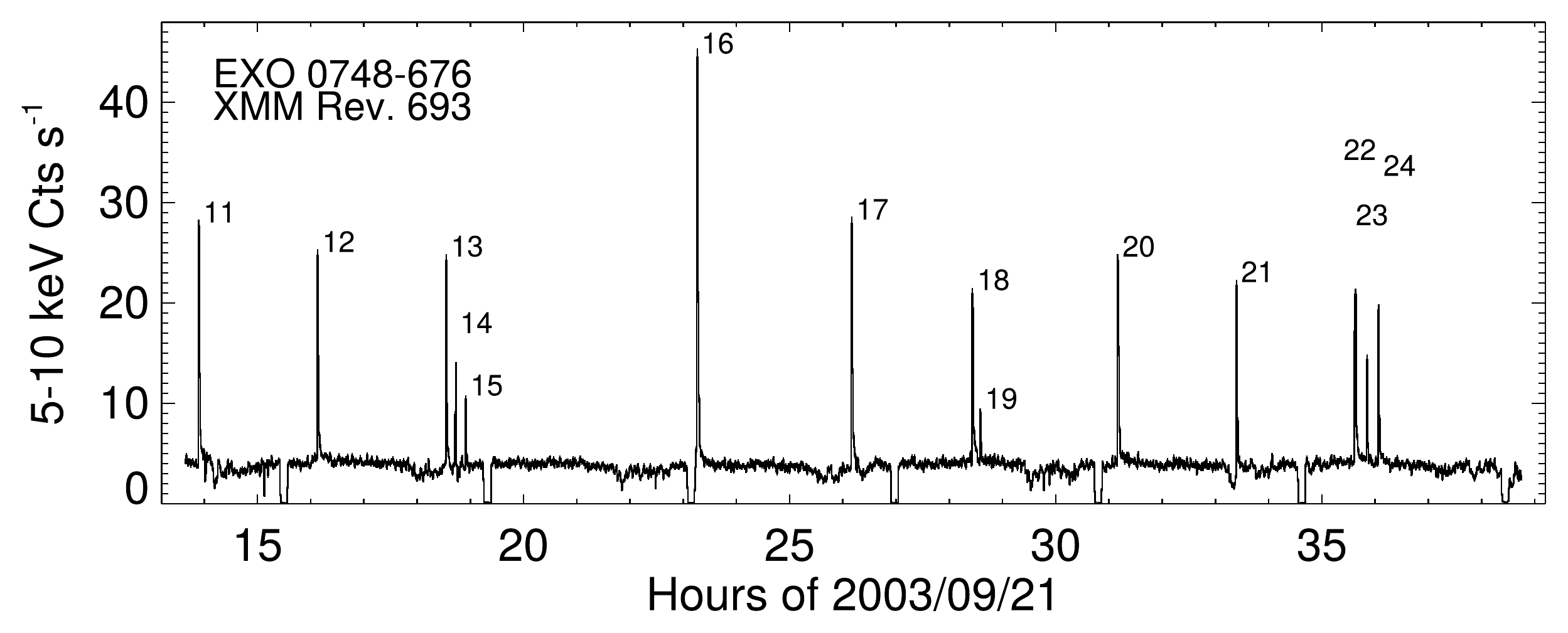}

\caption{Day-long light curve of EXO~0748$-$676 observed with {\it XMM-Newton} exhibits
double and triple events, where the bursts are separated by short recurrence times of 
$\sim 10$~minutes. The second and third bursts are typically
weaker than the first burst in an event, and the burst shape indicates that less
hydrogen is present in their fuel. Eclipses by the companion star periodically
obscure the source.
Reproduced with permission from \cite{boirin07a} \copyright\ ESO.
}
\label{fig:short}
\end{figure}

In the classical picture, a burst burns nearly all hydrogen and helium
in the neutron star envelope, and a completely new fuel layer must
be accreted before the next burst can ignite. Observations of bursts that recur within $\sim 5-20$~minutes suggest, however, a fraction of the fuel survives, as there is insufficient time to accrete a fresh layer \cite{1608:murakami80pasj,0748:gottwald87apj,fuji87}. Only recently have multi-zone simulations been able to show that a substantial fraction of the fuel may be left unburned after
a burst \cite{Keek2017}. Particularly if ignition occurs at a relatively
shallow depth (resulting in a weaker burst), over $\approx50\%$ of the
fuel remains unburned. Only the fuel near $y_{\mathrm{ign}}$ burns,
whereas the material at smaller depths is left-over. The recurrence
times of the bursts are reduced, since accretion needs only to replace
the burned fraction of the fuel column.

It is possible for the left-over fuel to ignite and produce a new
burst mere minutes after the previous burst (Fig.~\ref{fig:short}). If the fraction of unburned
fuel is large, this material resides relatively close to $y_{\mathrm{ign}}$.
Turbulent mixing can transport the unburned material down to this
depth, where it ignites a new burst. This secondary burst is typically weak,
with a lower peak flux and shorter duration than the previous burst,
because it is powered by a diluted mixture of fuel and ashes. Different
origins for the turbulent mixing have been proposed, such as rotationally
induced shear mixing \cite{fuji87} and post-burst convection
due to an inversion of the mean molecular weight \cite{ww84}\footnote{Irregular
bursting behaviour in early multi-zone models may have been an artefact of their
reduced nuclear reaction networks \cite{ww84,taam93}, as this behaviour is absent
in later work with large networks \cite{woos04}.}.
Recently, multi-zone simulations that include a full nuclear network
show that the turbulence is convection that is driven by an anti-correlation
between the temperature and the opacity in the ashes layer. As the
ashes cool after a burst, the opacity increases until convection switches
on. Convective mixing brings fresh fuel into the ashes layer, and
the recurrence time of the resulting burst is the cooling time of
the order of minutes. Because of the stochastic nature of convection,
sufficient fuel for a follow-up burst is mixed down only $\approx30\%$
of the time. Furthermore, after one follow-up burst, the mixing continues
and has the potential to produce another follow-up burst. This model
reproduces many of the observed features of such bursts  (Fig.~\ref{fig:short}): the follow-up
bursts have a lower peak luminosity and shorter duration than the first
burst, their recurrence time ranges from a few minutes to a few tens of
minutes, and several follow-up bursts can occur in so-called double, triple,
and quadruple events \cite{1608:murakami80pasj,0748:gottwald87apj,boirin07a,keek10}.

\subsection{Ignition Latitude}
\label{sec:latitude}

Up to now, we have only considered the radial dimension when locating the ignition point.
The latitude is also of importance: most bursting neutron stars spin rapidly 
(\S\ref{oscillations}), and consequently the effective surface gravity is reduced
near the equator compared to the polar regions (by a few percent for a typical burster \cite{slu02}).
The surface of a spinning star follows
an isobar (surface of constant pressure), and using the relation $y=-P/g$ from \S\ref{theory:accretion}, we see that
a smaller $g_\mathrm{eff}$ at the equator yields a larger $y$ than at the poles.
Furthermore, this implies that the {\em local} mass accretion rate is higher at the
equator. Because $y_\mathrm{ign}$ depends only weakly on $g_\mathrm{eff}$ (e.g., Equation 32 in \cite{bil98a}),
burst ignition preferentially occurs at the equator. 

The spreading of the flame on the surface of the star was modelled by \cite{slu02}.
The difference in speed between the spreading in the latitudinal and longitudinal 
directions (with the latter significantly faster at low latitudes) is thought to give rise to distinct,
observable differences in the shape of the burst rises, which is referred to as the
``convexity'' of the rise \cite{mw08}. However, a study of a large
sample of bursts observed by \xte\/ suggests that  bursts occurring when the source
is in the low (hard) spectral state typically ignite on the equator, while bursts
observed in the high (soft) spectral state ignite at higher latitudes \cite{mw08,zhang16}.
The obvious question is, why do the latter bursts not also ignite on the equator?

An attractive explanation is that in the high (soft) state, both H and He fuel burns
{\it steadily} on the equator, while unsteady burning (likely triggered by He) continues
in regions at higher and lower latitudes. This explanation would also account for the
infrequent, weak bursts that are typically observed in the high (soft) state (cf. with \S\ref{burstobsstatus}).
At present this scenario remains highly speculative, and it is computationally challenging
to test with multi-dimensional models (\S\ref{sec:multid_models}). Certainly, the
balance in pressure resulting from the decrease in effective gravity at the equator
gives a larger effective accretion rate per unit area there, but not sufficient to
stabilise the burning \cite{slu02}. 
Resolving this puzzle will likely require a substantial improvement in our understanding
of the global burst properties and perhaps also the influence of the accretion flow.

\section{The Burst Spectral Energy Distribution}

Most of the energy from nuclear burning is generated at least several
meters below the neutron star surface, and processed  before exiting the star from a cm-thick photosphere. The burst radiation is, therefore,
thermalized and leaves the burning layer 
as a blackbody spectrum. Scattering
off electrons and ions in the photosphere, however, modifies the continuum
spectrum and may introduce absorption lines and edges. Those departures
from a pure blackbody spectrum offer opportunities to measure the
neutron star mass and radius. In \S\ref{sec:interaction} we further discuss
how some fraction of the burst spectrum may be  reprocessed by the accretion environment.

\subsection{The Continuum Spectrum}
\label{sec:continuum}

\begin{figure}
\begin{center}
\includegraphics[width=\textwidth]{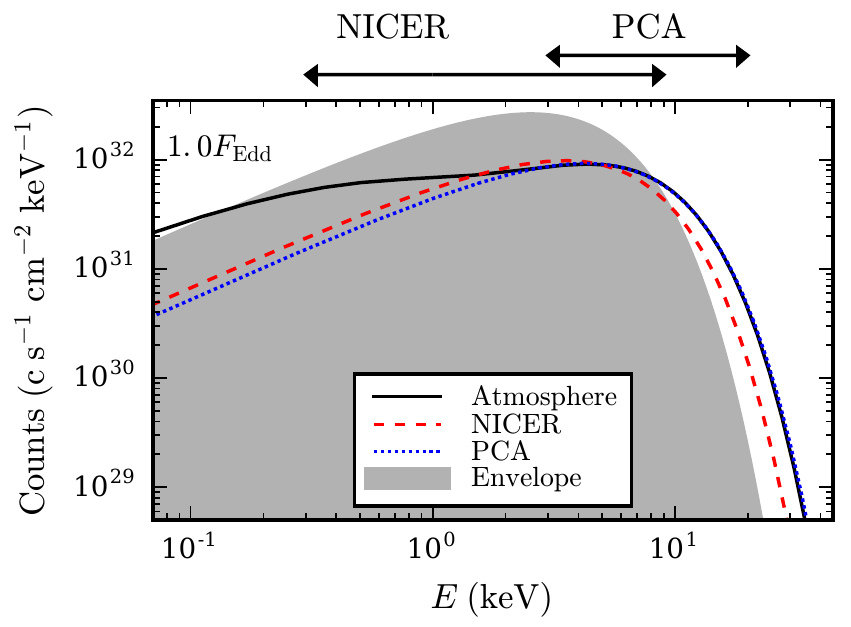}
\end{center}
\caption{\label{fig:example_spectra} Photon counts spectrum as a function of energy, $E$, for models of thermal burst emission (adapted from Keek et al. 2018 in prep). The top of the grey area marks the Planck spectrum in the neutron star envelope during a burst. Scattering in the atmosphere modifies the spectrum (solid line). Historically the observed spectrum is fit with a Planck model. Two such fits are shown for {\it NICER} and {\it RXTE}/PCA in the energy bands indicated at the top. The ratio of the observationally inferred temperature and that of the envelope spectrum is the colour correction factor.}
\end{figure}

The  majority of observed burst spectra are well described by a Planck
 (blackbody) distribution \cite{swank77}. The rise of the burst flux, set 
by a combination of the time scales of nuclear burning and flame spreading,
can be as fast as a millisecond, or 
as long as 
several seconds.
The measured blackbody temperature increases to a maximum value of
2--$3\,\mathrm{keV}$. In the burst tail, the decrease of the temperature is determined by the time scales
of cooling and of waiting points in the
{\sl rp}-process (\S\ref{sec:nuclear}),
and follows that of the burst luminosity,
typically lasting $\sim 10$--100~s. This cooling of the
neutron star atmosphere in the burst tail is often cited as prime
evidence of a thermonuclear flash (Type I burst) as opposed to an
accretion flare (Type II burst; \cite{lew93,Bagnoli2015}). 
A small subset of thermonuclear events exhibit only weak evidence of cooling.
In an analysis of the remarkable burst source and 11~Hz pulsar IGR~J17480$-$2446,
it has been shown that cooling may not be detectable when the ratio of peak
burst flux to persistent (accretion) flux is low, and the bursts correspondingly
have a low peak temperature \cite[]{linares11}.

Because
of the evolution of the spectral parameters, time resolved spectroscopy
has to be performed, with each spectrum being collected in a short time
interval. 
The short duration over which each spectrum is accumulated in typical bursts strongly limits the photon
counts and the detail present in a single spectrum. Only when considering
exceptionally long bursts \cite{Strohmayer2002} or large numbers
of bursts \cite{worpel13a} can deviations from a blackbody shape
be detected. Some deviations are due to changes in the persistent
spectrum induced by the burst (\S\ref{sec:interaction}), 
in addition to the modifications of the intrinsic spectrum discussed here.

Compton and free-free scattering of thermal emission in the neutron star atmosphere
modifies the burst spectrum (Fig.~\ref{fig:example_spectra}). Specifically, up-scattering boosts the energy
of a fraction of the soft photons, such that the ``colour temperature''
measured by fitting a blackbody is larger than the effective temperature
of the photosphere \cite{pa86}.
This effect is normally accounted for using a ``colour-correction'' factor, $f_{\rm c}$, giving the expected ratio of the (measured) colour temperature to the effective temperature. At fluxes close to Eddington, $f_{\rm c}$ is expected to be $\approx2$, dropping to $\approx1.4$ in the burst tail (e.g. \cite{suleimanov11a}).

Radiative transfer simulations
are employed to calculate all scattering processes in the atmosphere
taking into account the composition, although initial models were
not fully converged \cite{Madej2004,Majczyna2005}. Current state-of-the-art
spectral models are one-dimensional, assume hydrostatic and radiative
equilibrium, and have been calculated for a range of compositions
and surface gravities \cite{suleimanov11a,suleimanov12a,Nattila2015}. The models
have several limitations. They assume uniform emission across the
entire stellar surface, whereas the detection of burst oscillations
argues that this is not always the case (\S\ref{oscillations}).
Furthermore, the neutron star's rotation will cause Doppler broadening of
the spectrum \cite{Baubock2015}. The impact of these effects is,
however, likely small. More importantly, the assumed equilibria imply
that the models are not representative of the PRE phase, which will
require simulations that include both radiation transfer and hydrodynamics (cf. with \cite{pa86}). 
In a new approach, time-dependent stochastic (Monte Carlo) models are being
developed, which in time may overcome these limitations \cite{Medin2016}.

When observed in a restricted instrumental energy band, a neutron star atmosphere
spectrum is still well-approximated by a Planck function (Fig.~\ref{fig:example_spectra}). The inferred temperature
of the blackbody $T_{\mathrm{bb}}$ is expected to differ from the atmosphere's effective temperature $T_{\mathrm{eff}}$
by the colour-correction factor, $f_{\mathrm{c}}\equiv T_{\mathrm{bb}}/T_{\mathrm{eff}}$,
which is expected to vary as a function of the burst luminosity.
$f_{\mathrm{c}}$ has been calculated for assumed model atmospheres for application to \emph{RXTE}/PCA observations
\cite{suleimanov11a}, and can similarly be obtained for other instruments
using the model-predicted spectra. 

Aside from the temperature, the blackbody flux depends on the size
of the emitting area: the neutron star surface. It is typically assumed
that the whole surface radiates uniformly.\footnote{Emission from only part of the surface has been inferred for the 1999 superburst from 4U~1820$-$30 \cite{bml10}. This study ignored, however, disk reflection \cite{bs04}, which complicates the interpretation of superburst spectra \cite{keek14a,Keek2014sb2,keek15a}.} 
At the onset of the brightest
bursts, the 
flux may exceed the Eddington limit, 
causing photospheric radius expansion
and possibly also driving a wind from the neutron star surface \cite{Ebisuzaki1983,Quinn1985,Nobili1994,wbs06}.
During PRE the photospheric radius increases typically by up to a
factor $\sim10$, but even larger expansions have been observed (``superexpansion''; \S\ref{sub:intermediate-duration-bursts}).
Once the flux is
reduced below the Eddington limit, the photosphere settles back at
its original radius (the ``touchdown'' point). 
After touchdown, one expects the photospheric radius
to remain constant, but substantial evolution in the blackbody normalisation,
$K_{\mathrm{bb}}$, can be observed \cite{guver12a,zhang13}. We
discussed how deviations from a pure blackbody spectrum introduce
a colour correction factor for $T_{\mathrm{bb}}$. To preserve the
total blackbody flux $F\propto K_{\mathrm{bb}}T_{\mathrm{bb}}^{4}$,
this produces an equivalent correction to $K_{\mathrm{bb}}$: the
proportionality between the measured blackbody normalisation and the
colour correction is $K_{\mathrm{bb}}^{-1/4}\propto f_{\mathrm{c}}$ (see Fig.~\ref{fig:burstspec}).

The properties of the continuum burst spectrum have been employed to measure
the neutron star mass and radius (e.g.
\cite{Paradijs1979,Paradijs1986spectra,miller13,ml16}).
Two approaches have been used: the ``touchdown'' and the ``cooling-tail''
methods. The former method uses the flux at touchdown as a measure
of the Eddington limit, which depends on the neutron star mass,
in combination with measurements of the blackbody radius and (usually) the source distance 
\cite{Guver2010_1608,Guver2010}.
The radius is measured through the blackbody normalisation, $K_{\mathrm{bb}}\propto4\pi R^{2}$, 
assuming that $f_{\mathrm{c}}$ is constant.
It is possible
that the photosphere has not completely settled onto the neutron star at touchdown
\cite{Steiner2010}. Additionally, an accurate measure of $f_{\mathrm{c}}$
and the Eddington limit requires knowledge of the composition of the
photosphere as well as the source distance, which are generally known
only approximately.

The cooling-tail method uses the flux-dependent colour corrections from model atmosphere
spectra \cite{suleimanov11a,Suleimanov2011}. Because these models
are only valid at fluxes below the Eddington limit, only the tail
of a burst is considered. By measuring $K_{\mathrm{bb}}^{-1/4}$,
one observes how $f_{\mathrm{c}}$ changes as the burst flux decreases.
The atmosphere models predict $f_{\mathrm{c}}$ as a function of the
Eddington ratio and the neutron star mass and radius. Fitting these
$f_{\mathrm{c}}$ curves to the observed data allows for the neutron
star parameters to be measured. This method does not depend on prior
knowledge of the source distance or the atmosphere composition, as
the required information is encoded in the $f_{\mathrm{c}}$ curves. 

The radii obtained with the cooling-tail method are systematically
larger than with the touchdown method. Each method has a set of assumptions
that may be the source of this systematic uncertainty \cite{Steiner2010,guver12a,Guver2012Edd,poutanen14,Guver2016}.
For example, the inclusion of stellar rotation in the atmosphere models
may reduce the derived radius \cite{poutanen14}. Furthermore, each
method appears to work best for a different set of burst observations.
Bursts in the low (hard) flux state conform to the predictions of
the model atmospheres, whereas in the high (soft) flux state the bursts
do not exhibit the expected $f_{\mathrm{c}}-F$ relation \cite{kajava14}.
In the high flux state the accretion disk is expected to extend to
the neutron star surface, and a spreading layer may cover a substantial
part of the star \cite{Revnivtsev2013}, which reprocesses the burst
spectrum. Conversely, studies using the touchdown method have preferred
bursts that show only small deviations from a blackbody, rather than
those that follow the expected $f_{\mathrm{c}}-F$ relation \cite{guver12a}.
The source 4U~1608$-$522 exhibits bursts in both flux states, and an
analysis with both methods illustrates the differences in the obtained
results \cite{poutanen14}. Further details on the measurement of
mass and radius are presented in \cite{Miller2013}.

\subsection{Discrete Spectral Features}
\label{sec:features}

Before the thermalized burst spectrum leaves the neutron star, the
photons scatter off electrons in the photosphere. If the 
photosphere is incompletely ionized, 
discrete absorption features may appear in the spectrum.
The photon's energy may take the electron from one bound state to
another (bound-bound transition), which produces an absorption line
at that energy. Alternatively, the photon energy may exceed the ionization
energy required to unbind the electron (bound-free transition). This
creates an absorption edge that starts at the ionization energy and
continues towards higher energies. In the photosphere of an accreting
neutron star, hydrogen and helium are fully ionized. Metals, however,
may be only partially ionized and can produce lines and edges,
particularly in the extended photosphere driven by a strong radius-expansion burst (see \S\ref{sec:interaction}).
A range
of metals may produce these features, depending on the composition
of the photosphere \cite{wbs06}. Iron is typically one of
the most abundant heavy metals, and has relatively many transitions
to create discrete absorption features.

Detection of surface features has been a high observational priority for decades, due to the prospects of constraining the neutron star compactness via the gravitational redshift, measurable from comparing the observed and rest-frame energies of such features. Although there have been several notable claims for detections of features (see below), many have been unverified in subsequent observations and/or been shown to be unlikely to arise from the NS surface. There are a number of mechanisms which may act to remove elements that might give rise to features, from the upper layers of the photosphere. 

Gravitational settling is thought to rapidly separate accreted metals from the
photosphere on a timescale of $10^{-3}\ \mathrm{s}$ \cite{Brown2002,Peng2007}. The metals that can produce
lines and edges may, therefore, usually be absent from the neutron star surface.
Just a few meters below the surface, a plethora
of metals is created by nuclear burning during bursts. 
These elements may be revealed at the neutron star surface --- only during
the most powerful bursts --- because of
two processes. 
First, convection at the onset of the burst transports
the burning ashes towards the surface. Convection typically does not
reach the surface, but the more powerful the initial burning, the
closer to the surface the metal-rich ashes are convected. Secondly,
if the burst has PRE, a wind during the Eddington-limited phase may
blow off the upper atmosphere, facilitating the exposure of heavy
metals \cite{wbs06}. The appearance of metals in the photosphere
 can also give rise to a changing colour correction factor \cite{Kajava2016}.
In bursts where the fuel is helium rich, most of the burning happens
at the burst onset, and the Eddington limit is reached. Powerful helium
bursts are, therefore, the prime candidates for observing absorption
features during the PRE phase. For example, a weak absorption line
was detected during the PRE phase of an intermediate duration burst
from GRS~1741.9$-$2853 \cite{Barriere2015}.

The energy at which a line or edge is observed is reduced by the gravitational
redshift of the neutron star,
that also
provides a measure of the
star's compactness. 
Discrete features
are also expected to be affected by relativistic
Doppler broadening, since most bursting neutron stars rotate rapidly (e.g.\cite{baubock13}).
These
effects reduces the detectability of such features against the continuum,
increasing the difficulty of 
constraining the
equation of state. 

The studies of EXO~0748$-$676 provide an illustrative example.
The source was a calibration target for \emph{XMM-Newton}, and consequently
a large number of bursts were observed with the Reflection Grating
Spectrometer. Stacking the spectra of a series of bursts revealed
a tentative detection of an iron absorption line \cite{Cottam2002,Rauch2008}.
The inferred gravitational redshift of the line was employed to constrain the
mass and radius of this neutron star \cite{Ozel2006}, favoring a
soft equation of state. These analyses assumed a low neutron star
spin frequency, supported by the detection of a weak burst oscillation at $45\,\mathrm{Hz}$ \cite{Villarreal2004}. Later,
a much stronger burst oscillation was deteted in two bursts, at a significantly 
higher frequency: $552\,\mathrm{Hz}$ \cite{gal10a}. 
If the later detection instead corresponded to the neutron star spin in this system,
the expected Doppler broadening
was, therefore, substantially larger than assumed, casting
doubt on the line origin. Furthermore, a larger sample of bursts
observed with \emph{XMM-Newton} \cite{Cottam2008} and \emph{Chandra}
\cite{keek10} do not exhibit the line feature. The status of the line detection
from EXO~0748$-$676 remains, therefore, uncertain. 
However,
the potential for absorption features from the neutron
star surface to constrain the neutron star equation of state
remains high.

Because a wider range of energies is absorbed to produce an edge,
edges typically have a larger equivalent width than lines \cite{wbs06},
and consequently should be easier to detect. The strongest absorption
features associated with the surface of neutron stars are nickel edges
observed in the PRE phase of three superexpansion bursts from 4U~0614+091
and 4U~1722$-$30 \cite{Zand2010}. These detections were made with
the \xte/PCA, which has modest spectral resolution ($\Delta E/E\approx0.17$ at 6~keV, where $\Delta E$ is the full-width at half maximum; \cite[]{xtecal06}). Improved constraints
on the neutron star equation of state will result from future observations
of edges in helium-rich bursts performed with high-resolution spectrometers,
but it has proven challenging to schedule such observations of relatively
infrequent helium flashes.

\section{Interaction with the Accretion Environment}
\label{sec:interaction}
The dynamical influence of thermonuclear bursts on the accretion disk has generally been studied in the context of radius-expansion bursts, which exceed the Eddington limit at the neutron star surface, and may be expected to drive winds into and across the disk. 
More broadly, three principal effects have been discussed in the context of interactions between bursts and the accretion disk
\cite[e.g.][]{ballantyne05}: structural changes to the disk (and/or corona) resulting from burst-induced X-ray heating (or cooling); radiatively (or thermally) driven outflows; and inflow due to Poynting-Robertson drag. In addition, the observed spectrum may be modified by interaction with, and reflection from, the accretion disk.

The evidence for structural changes comes from some observations of long-duration bursts (see also \S\ref{sec:long-duration-bursts}). In some of these events we observe strong flux variability late in the tail of the burst \cite[]{zand11a,degenaar13a} (Fig.~\ref{fig:intermediate_duration_burst}).
This variability has been suggested to arise from Thompson scattering in the settling ``fragments'' of disk material, following disruption by an extremely energetic radius-expansion burst. 
Decreases in the intensity of high-energy (30--50~keV) photons beginning a few seconds after the burst onset have also been observed in a few sources \cite[]{mc03,ji14}. These photons are thought to arise from Compton scattering in a hot corona, and it is expected that the sudden rise in (relatively) cool photons will result in a transient change in the coronal structure. While the principal effect predicted is additional cooling of the corona, the estimated change in the cooling rates is in the other direction, i.e. predicting {\em reduced} cooling \cite{ji15}. 
Such explanations also seem at odds with the evidence supporting the role of magnetic fields energizing coronae in other accreting sources (e.g. \cite{mf01}).
Other possible mechanisms which might instead give the observed behaviour are related to the attenuation of the inner disk in response to the burst (see below), possibly mediated via the magnetic field threading the disk  \cite[]{ji14}. 

Irradiation of the accretion disk may drive a wind from its surface. The prospects of this phenomenon have been discussed in
the context of spectral features observed from an intermediate duration burst of IGR~J17062$-$6143 \cite{degenaar13a}.
Additionally, such powerful bursts may eject a shell from the atmosphere of the neutron star in a superexpansion phase,
and enrich the surroundings in metal-rich burst ashes \cite[]{wbs06}. This merchanism has been argued to be the source of discrete
spectral features detected in intermediate duration bursts (\cite[]{zand10a}; see also \S\ref{sec:features}).

The effects of Poynting-Robertson drag arising from the increase in X-ray flux during the burst have been predicted to lead to an increase in the accretion rate (and hence the persistent flux component) during the burst \cite[]{walker92} (see also \cite{Lamb1995}). 
This effect is potentially at odds with 
the conventional approach for time-resolved spectroscopic analysis,
which involves subtracting off the pre-burst emission (usually incorporating both the persistent emission and instrumental background). This approach implicitly assumes that the burst and persistent component are independent, even late in the tail of the burst, which cannot be completely true (e.g. \cite{kuul02a}).
Nevertheless, the net burst spectrum is then usually well-fit with a blackbody (e.g. \cite{swank77}). 

\begin{figure}[ht]
\begin{center}
\includegraphics[width=0.7\textwidth]{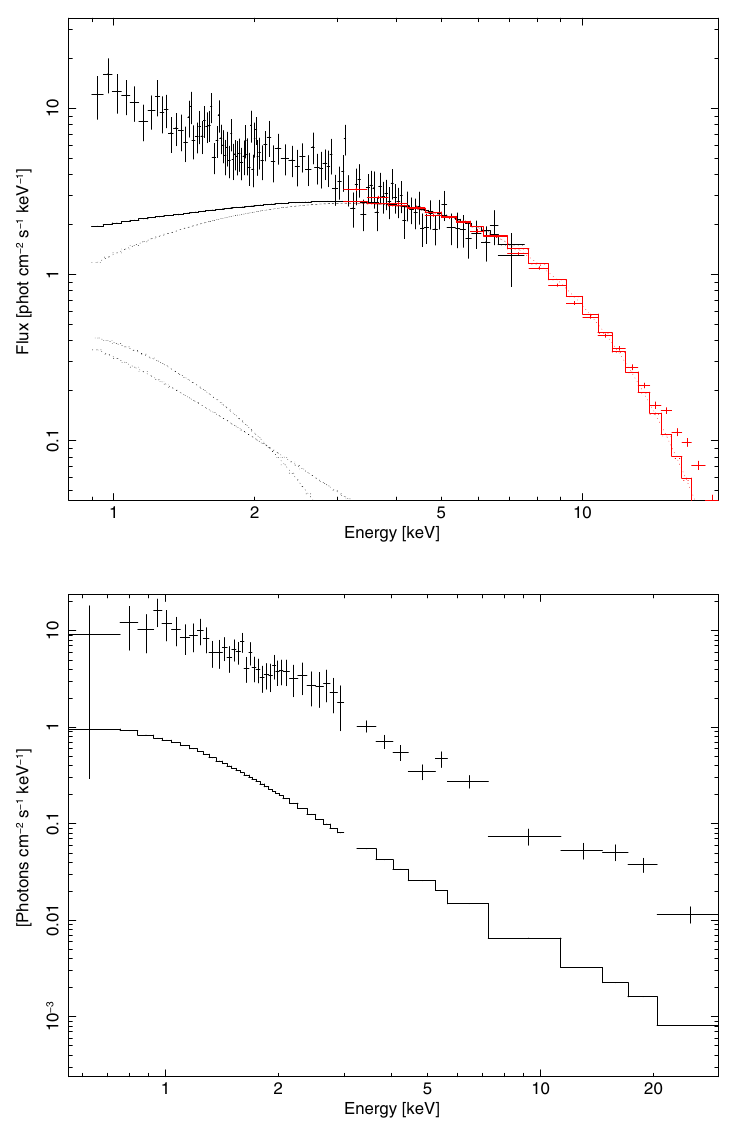}
\end{center}
\caption{X-ray spectra during a bright burst from the millisecond pulsar SAX~J1808.4$-$3658, observed simultaneously with \xte\/ and \chandra, illustrating the apparent increase in the persistent flux level during the burst.
The top panel shows the \chandra\/ (black crosses) and \xte\/ (red crosses) spectrum, integrated over an interval 2 to 5 s after burst onset. 
The solid  curve shows the best-fit blackbody spectrum, including the
model components for the pre-burst (persistent) radiation. The model is a poor fit (reduced $\chi^2_\nu = 3.505$ for $\nu = 268$ degrees of freedom), but is dramatically improved if the normalization  of the persistent component is also free to vary ($\chi^2_\nu = 0.963$ for $\nu = 267$). 
In the bottom panel the residual spectrum is plotted after subtraction of the fitted black body model (crosses). The model for the non-burst spectrum is plotted as a histogram. For clarity \chandra\/ data is plotted up to 3 keV and \xte\/ data beyond 3 keV. 
Note the similarity in shape of the residual  and the persistent spectrum. 
Reproduced with permission from \cite{zand13a} \copyright\ ESO.
\label{fig:sax1808burst} }
\end{figure}

The burst sample accumulated by \xte\/ over its mission offers 
the most
stringent test of this conventional analysis approach. A study of 332 PRE bursts observed from 40 sources found significant evidence for an increase in the persistent contribution to the flux early in the bursts \cite{worpel13a}. This increase, up to a factor of 20, was interpreted as the expected increase in the accretion rate in response to Poynting-Robertson drag. It is somewhat surprising that this effect is seen in PRE bursts, since the luminosity of those bursts should exceed the Eddington limit at the peak, temporarily {\it halting} accretion. The effect has also been confirmed in an even larger sample of 1759 non-PRE bursts from 56 sources \cite{worpel15}, although at with smaller increase factors.
Additionally, a bright burst from SAX~J1808.4$-$3658 observed simultaneously with {\it Chandra}\/ and \xte\/ finds even more compelling evidence for increases in the accretion component during the burst \cite{zand13a}, from the much broader band-pass ($\approx0.5$--30~keV) available for the two instruments combined (Fig.~\ref{fig:sax1808burst}).

Unfortunately the available data is insufficient to test for possible changes in the {\it shape} of the persistent spectrum accompanying the burst-induced increase in accretion rate,
which might be expected in parallel to the spectral shape variations that are seen on longer timescales (see e.g. \S\ref{burstobsstatus}).
Furthermore, while the modified analysis approach improves the combined distribution of fit statistics for the ensemble, it does not yet achieve the expected distribution for a well-fitting model, suggesting that the net burst spectrum still deviates significantly from a blackbody. As we have discussed, these deviations include the recombination edge features observed in very energetic bursts (e.g. \cite{zand10a}; see also  \S\ref{sec:long-duration-bursts} and \S\ref{sec:features}). The residuals also appear inconsistent with the more modest deviations predicted by spectral models (see \S\ref{sec:continuum}).
The question of the true shape of the burst spectrum over the full range of burst types remains open, with the available data not sufficiently sensitive to clearly measure deviations from a blackbody. However, with the analysis of the \xte\/ sample, the key contributions of variations in the persistent contribution, discrete features, as well as the effect of reflection from the accretion disk (\S\ref{sec:reflection}) are now well-established.

\subsection{Reflection by the Accretion Disk}
\label{sec:reflection}

\begin{figure}
\includegraphics[width=\textwidth]{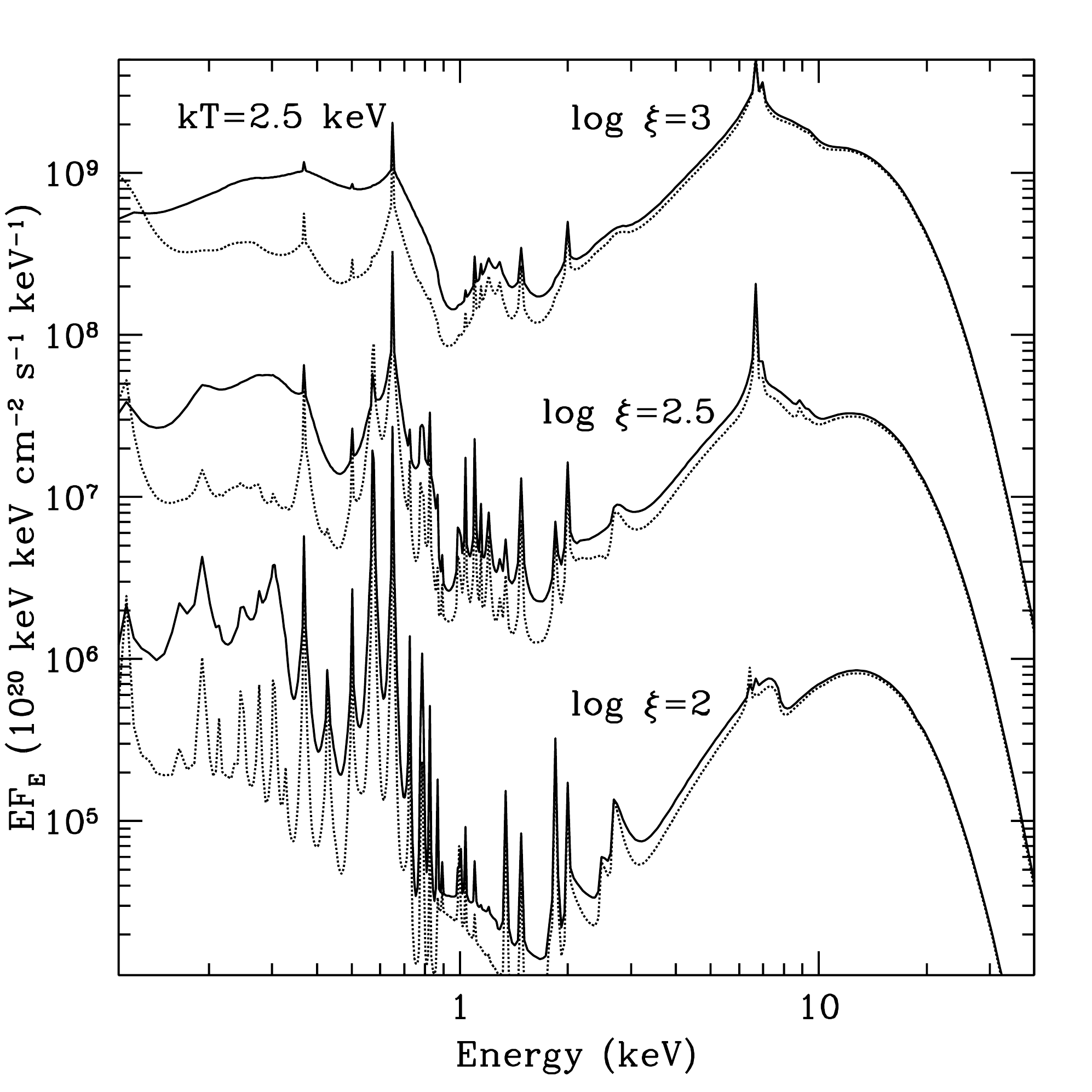}

\caption{Models of burst reflection spectra  as a function of energy (from \cite{Ballantyne2004models}).
For an irradiating blackbody spectrum with $\mathrm{k}T=2.5$~keV, the resulting
reflection spectra are shown for three values of the disk ionization parameter
$\xi$. Especially below $\approx3$~keV, the reflection spectra depend on the density
of the inner disk ($n=10^{18}\ \mathrm{cm}^{-3}$ for solid lines and $n=10^{15}\ \mathrm{cm}^{-3}$
for dotted lines). Observed reflection spectra are typically modified by relativistic
Doppler broadening.
}
\label{fig:reflection}

\end{figure}

A substantial fraction of the X-rays emitted by the bursting neutron star fall
on the accretion disk \cite{lapidus85mnras,fujimoto88apj}. For a thin disk this 
fraction is about $25\%$ of the total burst emission, and most flux falls on 
the inner few tens of kilometers closest to the star. The radiation undergoes 
diffusive scattering off the disk's photosphere, which is referred to as X-ray
``reflection.'' Depending on the inclination angle of the disk with respect to
the line of sight, a substantial part of the observed burst flux may have
arisen from photons
reflected by the disk. The reflection fraction is defined as the ratio of the
reflected and the directly observed burst flux components. For a thin disk, reflection
fractions of up to approximately $0.5$ are expected. 

If the disk has a different shape,
such as a steeply increasing height, the reflection fraction can potentially exceed
unity: most of the burst signal is observed in reflection \cite{he16}. A simple
analogy is a candle in a teacup: when viewed under certain angles, the flame of the
candle is not directly visible, but one still sees its light scatter off the inner
sides of the cup.

When the burst scatters off the disk, the X-rays are reprocessed, and their
spectrum is modified \cite{Ballantyne2004models}. Above $3$~keV, the reflection spectrum is shaped like the
illuminating burst spectrum with the addition of a fluorescent iron emission line near $6.4$~keV
and an iron absorption edge near $9$~keV (Fig.~\ref{fig:reflection}). The shape of the line depends on the ionization
state of iron in the disk, quantified by the ionization parameter $\xi =4\pi F n^{-1}$,
with $F$ the illuminating flux measured at the disk and $n$ its number density.
Reflection further contributes a number of emission features below $3$~keV as well
as a free-free continuum. The strength of these features depends on the density and
composition of the reflecting material (Fig.~\ref{fig:reflection}).

The reflection spectrum undergoes relativistic Doppler broadening
from the motion of gas in the disk,
which is stronger
if the inner disk is closer to the neutron star (cf. with \S\ref{burstobsstatus}). A reflection spectrum, therefore,
contains information on the location of the inner disk, its ionization state, composition,
and density, and the geometry of the disk. Current reflection models cover only part
of this large parameter space, making accurate interpretation of the spectra challenging.
Thus far, reflection has been detected in two superbursts 
\cite{stroh02,bs04,keek14a,keek14b,keek15a} and one intermediate
duration burst \cite{Keek2016igr1706}. The reflection signals hint at a strong impact
of these bursts on their environment, showing the evolution of the disk inner radius
and its ionization parameter.

\subsection{Anisotropic Emission}
\label{sec:anisotropy}

An interesting consequence of reflection combined with obscuration of the neutron star by the
disk is that the observed burst flux depends on the inclination angle of the disk
with respect to the line of sight. Both the burst flux from the neutron star and the
flux from the disk are anisotropic. If the inclination angle is known, the anisotropy
factors $\xi$ can be calculated, which are defined as $L=4\pi d^2\xi F$, with $L$ and $F$
the intrinsic luminosity and the observed flux, respectively, of either the star or
the disk, and with source distance $d$ \cite{lapidus85mnras,fujimoto88apj,he16}.

Unfortunately, the inclination angle of most bursting systems is poorly constrained (e.g. \cite{gal16a}). Moreover,
most predictions for $\xi$ assume that the accretion disk is thin and extends down to the
surface of the neutron star \cite{lapidus85mnras,fujimoto88apj}. At lower mass accretion rates, however,
the disk is truncated at some distance from the star. Also, different disk shapes will alter the anisotropy $\xi$ 
\cite{he16}. Lack of information on the disk geometry hampers accurate predictions of $\xi$.
The observation of reflection fractions in excess of $\gtrsim 0.5$ (the maximum for thin disks)
indicates a concave geometry. For the two superbursts observed with \xte/PCA large reflection
fractions in excess of unity were inferred \cite{bs04,keek14b}, suggesting such a disk shape.
However, because of the limited data quality alternate interpretations are possible that are
consistent with a thin disk \cite{keek15a}.

The anisotropy has important consequences for a range of measured quantities, from the burst
peak flux and fluence, to the mass accretion rate derived from the persistent flux. It impacts
distance measurements that use the peak flux of Eddington-limited bursts. Different values of
the Eddington flux have been derived from bursts of the same source \cite{gal06a},
but the effect of variations in $\xi$ in different accretion states has not been investigated
yet. Radius measurements are impacted as well, as they are often derived from the normalization
of the spectrum. Unfortunately, the effects of anisotropy are often neglected, because the 
system inclination
is difficult to measure accurately.

\section{Burst oscillations and the neutron star spin}
\label{oscillations}

The evolutionary role for accretion-powered neutron stars in low-mass binaries 
as precursors to rotation-powered millisecond pulsars
\cite{alpar82} has been well established with the discovery of several
types of coherent millisecond oscillations, by which the neutron-star spin
frequencies may be measured. Most compelling has been the discovery of
persistent X-ray pulsations at frequencies typically in the range
200--600~Hz (e.g. \cite{pw12}).
Those sources also confirmed the earlier discoveries of burst oscillations as an alternative tracer of the neutron star spin \cite{chak03a}.

Burst oscillations are coherent, periodic variations in the X-ray intensity of bursting sources that are observed  during a thermonuclear burst (see \cite[]{watts12a} for a review).
These oscillations were first detected in the well-studied burst source 4U~1728$-$34 by {\it RXTE}\/ shortly after its launch, in 1996 \cite{stroh96}. 
To date oscillations have been detected in 17 sources, with tentative (low significance, or single-burst detections, or both) claims for a further 9 sources.
The oscillations are detected at a typical fractional amplitude of 0.05--0.2, and are within a few Hz of a frequency value characteristic for each source. 
The oscillation frequency tends to drift upwards by 1--2~Hz while it is present.
For those sources which also show pulsations in the persistent (accretion-powered) X-ray emission, the frequencies of the persistent pulsations and burst oscillations are within a few Hz of each other, confirming the burst oscillation frequency as (roughly) the neutron star spin \cite[]{chak03a}. 
Oscillations are not observed in every burst, but instead are preferentially seen at high accretion rates \cite[]{muno04a,ootes17a}, when the spectrum is in the ``soft'' state (see \S\ref{burstobsstatus}). 
Even in those bursts where they are present, the oscillations are not found throughout the burst, instead tending to occur most often in the burst rise, and less frequently through to the burst tail \cite{bcatalog}.

The mechanism giving rise to the oscillations remains unknown; candidates include the spreading hotspot model, the related ``cooling wake'' that follows, and models involving surface modes of various kinds. The rise in fractional burst oscillation amplitude in the early phases of the burst generally supports the hotspot model \cite{stroh97b,chakraborty14},  although once the burning has spread to the entire surface it is difficult to understand how oscillations can persist. If the cooling timescale is similar to the spreading timescale, it seems possible that the temperature contrast can also produce oscillations in the tail \cite{ms14}, although such models have difficulty achieving the high amplitudes that are observed.
Models invoking rotation modes \cite{heyl04} suffer from difficulties reproducing the observed frequency drifts \cite{bl08}.

In the absence of certainty as to the mechanism, much effort has been focussed on understanding why oscillations are not seen in every burst. Studies of the assembled \xte\/ sample have confirmed that oscillations preferentially occur at higher accretion rates \cite{ootes17a}.
There is some evidence \cite{zhang13,zhang16} linking the presence of oscillations in the burst tails with the characteristic evolution of the blackbody normalisation (proportional to the emitting area, up to a factor of the colour correction; cf. with Fig.~\ref{fig:burstspec}). However, since the spectral evolution is also correlated with the persistent spectral state (e.g. \cite{kajava14,poutanen14}, the causal relationship between these phenomena remains unclear.

At the same time, theoretical studies have revisited the microphysics of the burning taking place in bursts. 
Simulations of flame propagation for helium detonations, with possible application to intermediate-duration bursts (see \S\ref{sub:intermediate-duration-bursts}) have been performed \cite{Zingale2001,Simonenko2012}; however,  most bursts are likely deflagrations rather than  detonations.
Other simulations have modelled the propagation of the burning front (e.g. \cite{cavecchi13}), and the influence of the latitude of the ignition point and the neutron star spin. Spreading is understood to occur at different speeds in the latitudinal and longitudinal directions, and even at different speeds towards and away from the equator \cite{slu02,cavecchi15}.
The discovery of Terzan~5~X-2, a burster with a spin rate of 11~Hz (more than an order of magnitude slower than the next fastest burster) has provided an additional avenue to such studies \cite{cavecchi11}. 

Along with accretion-powered pulsations, burst oscillations have held the promise for constraining neutron star mass and radius, via measurements of the Doppler shift of the burst spectrum. However, the most recent studies remain inconclusive \cite{artigue13}, suggesting that significantly higher signal-to-noise is required, as might be provided by a next-generation mission such as the 
Enhanced X-ray Timing and Polarization mission ({\it eXTP}; \cite{extp16}), currently in phase A study with a launch expected from 2025 or beyond.
Another proposed mission offering effective area significantly in excess of the {\it RXTE}/PCA is the 
Spectroscopic Time-Resolving Observatory for Broadband Energy X-rays ({\it STROBE-X}; \cite{strobex17})

\section{mHz oscillations and marginally stable burning}

\label{sec:mHzQPOs}

Quasi-periodic oscillations (QPOs) at mHz frequencies have been detected from
$\approx5$\% 
of bursting sources. The behaviour of the mHz QPOs 
is related to the occurrence of bursts, and thus the QPOs are thought to result
from an oscillatory burning mode, as expected from marginally stable burning
(regime VI in Table \ref{tab:burning_regimes}). Numerical simulations of this burning regime
broadly reproduce the observed properties of mHz QPOs, although
some of the details do not match. The discrepancies are principally related to the  difficulty
in matching theoretically-predicted burning regimes to the observed mass accretion rates,
and mHz QPOs may provide important insight in how to improve theoretical
expectations. Here we review the observations of the mHz QPOs and
the theory of marginally stable burning, and discuss to what extent
they match.

\subsection{Observations of mHz QPOs}

\begin{figure}
\includegraphics[width=\textwidth]{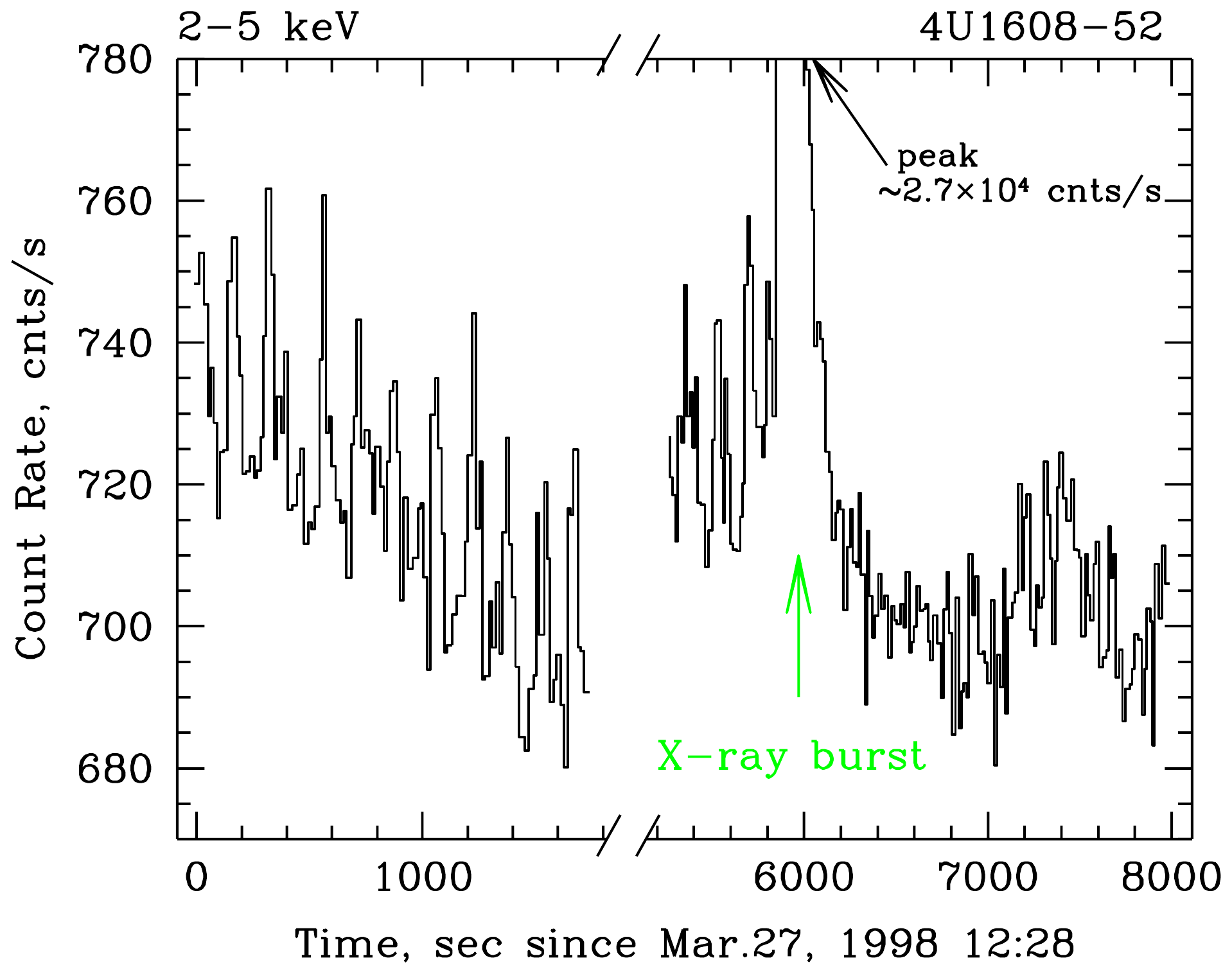}

\caption{mHz QPOs are observable in the light curve of 4U~1608$-$522 up 
to the X-ray burst around $6000$~s. The amplitude
of the oscillations is much smaller than the burst, and the burst peak is
not visible in this zoomed-in figure. After the burst
the oscillations are no longer detected, and the visible variability is Poisson
noise. It suggests that the mHz oscillations originate from the same location
as the burst, and are likewise produced by nuclear burning in the neutron
star envelope (regime VI in Table~\ref{tab:burning_regimes}).
Reproduced with permission from \cite{Revnivtsev2001} \copyright\ ESO.}
\label{fig:mhzqpos}
\end{figure}

QPOs at frequencies of 5--10~mHz with fractional amplitudes of $\approx 2\%$ 
have been detected in observations of $5$ bursting
sources: 4U~1636$-$536, Aql~X-1, 4U~1608$-$52 \cite{Revnivtsev2001} (Fig.~\ref{fig:mhzqpos}),
Terzan~5~X-2 \cite{Linares2011}, and 4U~1323$-$619 \cite{Strohmayer2012}.
All sources likely accrete hydrogen-rich material, judging from the
properties of the Type I bursts that they exhibited. The power spectrum
reveals an oscillation mode with a period of a few minutes, and the
oscillations can often be seen by eye in the X-ray light curve. The QPOs
occur in a narrow range of persistent flux (or mass accretion rate)
for each source \cite{Linares2011,Lyu2015}. For 4U~1636--536 it
was shown that the frequency evolves with time, and when it drops
below a certain value, an X-ray burst occurs \cite{Altamirano2008,Lyu2014,Lyu2015}.
This correlation is a strong indication that the oscillatory behaviour is related
to the nuclear burning processes in the neutron star envelope. After
the burst, the mHz QPO is absent for a while, and it returns to repeat
the frequency evolution before the occurrence of the next burst. There
is no apparent correlation between the frequency evolution and the
persistent flux. Furthermore, no (strong) correlation has been found
between the frequency and the photospheric temperature \cite{Lyu2014,Lyu2015}.
However, the limits are still consistent with predictions from multi-zone
models. Phase-resolved spectroscopy has been performed for one case, and it suggests
that the oscillations are produced by a hot spot of constant temperature that
is periodically changing in size \cite{Stiele2016}. This is at odds with the common
theoretical interpretation, but it remains unclear whether it signals a
failing of the theory or is the result of the limited quality of the data.

The behaviour that we described was found for all mentioned sources,
except Terzan~5~X-2. The behaviour of the bursts and mHz QPOs of this
source has been unique among the known bursters. This transient source
went into a month-long outburst in 2011, in which the accretion rate steadily
increased and subsequently decreased until the source was once again
quiescent. During the outburst rise, the burst rate steadily increased until
the recurrence time was a mere 3 minutes, and the light curve 
exhibited oscillations.
This interval was followed by an absence of bursts or oscillations
at the highest persistent flux. As the flux gradually returned to
quiescence, the reverse was observed: oscillations reappeared, followed
by bursts, whose recurrence time increased as the flux decreased. The behaviour of 
Terzan~5~X-2 is different from the other four sources, in which bursts and QPOs alternated
while the persistent flux remained approximately constant
at a value corresponding to a much lower accretion rate. Furthermore, for those
sources there is a large difference between the burst recurrence time
of several hours and the oscillation period of a few minutes.
For Terzan~5~X-2 there was no clear transition at which bursts ceased and mHz QPOs appeared.

\subsection{Theoretical interpretation: marginally stable burning}
\label{subsec:msb}

From the apparent interaction between the mHz QPOs and Type I bursts,
as well as from the energetics of the observed oscillations compared
to the persistent flux, it was argued that the mHz QPOs could be produced
by a nuclear burning mode \cite{Revnivtsev2001,Altamirano2008}. Oscillatory
burning modes are predicted to be present at the transition of stable
and unstable burning \cite{Paczynski1983}, because of competition
between heating and cooling processes. Stable burning results from
an equilibrium between cooling processes and heating by nuclear burning,
taking into account the rate at which fresh fuel is accreted. Conversely,
bursts appear after a period of accretion during which cooling was
sufficiently efficient to prevent most nuclear burning. Once the bottom
of the fuel column is compressed to reach a sufficiently high temperature
and density, the nuclear burning rate runs away unstably, quickly
overtaking the cooling rate and producing a burst. At the transition
between the stable and unstable burning modes, nuclear burning is
``marginally'' stable: the nuclear burning rate starts to rise,
but the cooling rate quickly catches up and prevents a runaway. This
alternates to produce an oscillatory burning mode with a period, $P$,
that is the geometric mean of the cooling (thermal) timescale, $t_{\mathrm{therm}}$,
and the accretion timescale (time to replenish the burned fuel), $t_{\mathrm{acc}}$:
$P\simeq\sqrt{t_{\mathrm{therm}}t_{\mathrm{acc}}}$ \cite{Heger2005}.
This picture is confirmed by multi-zone numerical simulations. The
marginally stable burning mode is reproduced in a narrow range of
mass accretion rates at the stability transition. For current standard
theory, this transition takes place at the Eddington limit, whereas
mHz QPOs are typically observed at a ten times lower mass accretion
rate. 

The mass accretion rate of the stability transition ($\dot{M}_{\mathrm{st}}$) depends on
the conditions in the neutron star atmosphere and potentially the burning lattitude.
These conditions include the effective gravity in the neutron star envelope and the composition
of the accreted material \cite{Heger2005,keek14b}. Equally important
are the details of the processes that take place in the atmosphere.
In particular, 
an increased heat flux into the atmosphere substantially
lowers $\dot{M}_{\mathrm{st}}$ \cite{keek09}. Similarly, turbulent
mixing induced by rotation or a magnetic field lowers $\dot{M}_{\mathrm{st}}$
by reducing the time it takes to bring fresh fuel to the burning depth
\cite{Piro2007,keek09}. Furthermore, several nuclear reactions
have been identified that play a key role in setting the stability
of the burning processes. The CNO breakout reaction $\mathrm{^{15}O(\alpha,\gamma)^{19}Ne}$
is one of the most important in this respect \cite{Fisker2006,Fisker2007,Davids2011},
but its reaction rate is poorly constrained by nuclear experiments,
leading to a substantial uncertainty in $\dot{M}_{\mathrm{st}}$ \cite{keek14b}.
Competition of this reaction with another breakout reaction, $\mathrm{^{18}Ne(\alpha,p)^{21}Na}$,
has been shown to increase the range of $\dot{M}$ where burning is
marginally stable \cite{keek14b}.

The oscillation period depends on $t_{\mathrm{acc}}$, and is therefore
expected to be a function of $\dot{M}_{\mathrm{st}}$. Similarly,
the waveform and the amplitude of the oscillations depend on $\dot{M}_{\mathrm{st}}$.
Additionally, changes are expected as a function of the ``distance''
from the stability transition. For example, at a mass accretion rate
close to $\dot{M}_{\mathrm{st}}$, the amplitude is smaller, the waveform
is more symmetric, and the frequency is higher \cite{keek14b}.

Marginally stable burning is an elegant explanation for mHz QPOs,
as it is expected to occur in a narrow range of mass accretion rate,
broadly matching observations.
The
predicted periods are consistent with the observed values of a few
minutes. However, if the observed accretion timescale is considered,
the predictions for the period are actually larger. Furthermore, marginally
stable burning takes place at the transition from bursts to stable
burning; this appears to be the behaviour observed from Terzan~5~X-2,
but not 
for the other sources,
where bursts and mHz QPOs alternate, and the transition to stable
burning takes place at higher persistent flux \cite{Keek2008}. Instead,
these sources 
more closely resemble 
models that include a slowly decreasing
heat flux into the neutron star envelope. Those simulations display
oscillations with an evolving frequency up to the ignition of a burst
\cite{keek09}. This suggests that the observed mHz QPOs are indicative
of cooling of the atmosphere, possibly heated by a preceding burst
\cite{Lyu2015}. However, multi-zone models that self-consistently
include the heating and cooling of deeper layers, do not exhibit alternating
oscillations and bursts. Therefore, while marginally stable burning
is the most likely process to power mHz QPOs, the circumstances under
which it is observed are challenging to explain with current theory.

\section{Burst duration and fuel composition}
\label{sec:long-duration-bursts}

The  majority of observed bursts have durations
of $\sim10\,\mathrm{s}$ to $\sim100\,\mathrm{s}$. A typical burst
ignites at a depth of $y\simeq10^{8}\,\mathrm{g\,}\mathrm{cm^{-2}}$,
where the thermal timescale is of the order of $10\,\mathrm{s}$.
If all fuel burns quickly after ignition, this cooling timescale sets
the burst duration. If the fuel is hydrogen-rich, however, nuclear
burning via the {\sl rp}-process can prolong the burst for up to
$\approx100\,\mathrm{s}$ \cite[e.g.,][]{schatz01}. 

Historically the burst timescale has been measured in a number of ways that sometimes make it difficult to compare different analyses. 
Measurements that rely on the length of the interval during which the burst emission exceeds some factor of the persistent noise level, are subject to variations due to different instrumental sensitivities.
Burst decays have been fitted with single (or sometimes multiple) exponential decay segments, with the decay timescale quoted as a measure of the burst duration (e.g., \cite{bcatalog}). In high signal-to-noise lightcurves, such models generally offer poor fits to the decay, and in fact more recent analysis favours power-law or more complex combinations \cite{zand14a,zand17b}.
A third, oft-quoted measure is the ratio of the burst fluence to the peak flux, i.e. $\tau=E_b/F_{\rm peak}$ (e.g., \cite{vppl88}). However, this quantity does not offer an unambiguous mapping to the burst fuel composition. For example, mixed H/He bursts with long {\sl rp}-process tails may have similar $\tau$-values as long, intense PRE bursts arising from ignition of a deep pile of pure He.

In rare cases bursts have
much longer durations, of minutes or tens of minutes (intermediate duration
bursts) to many hours (so-called ``superbursts''). These durations cannot
be explained by prolonged burning, but 
are comparable to 
the cooling timescales
of deeper layers. As it takes a longer time to accumulate a larger
fuel layer, these bursts have long recurrence times of months to years.
Observations are, therefore, rare, and have mostly been performed with wide-field
or all-sky instruments, which yield data of modest quality. 

The long bursts reach similar peak fluxes and photospheric temperatures
as regular bursts. The intermediate duration bursts all reach the
Eddington limit, whereas most superbursts have a lower peak flux.
From an observer's perspective, the main discriminating characteristic
is the long duration and correspondingly large fluence
of these events (Table~\ref{tab:durations}).

Long bursts are of particular interest, because they probe
deeper regions of the neutron star envelope down to the outer crust.
Furthermore, long durations allow for more detailed spectra to be
obtained, displaying interesting behaviour such as interaction with the
accretion environment (\S\ref{sec:interaction}).

\begin{table}

\caption{\label{tab:durations}Typical properties of Type I bursts with different
durations}

    \begin{centering}

\begin{tabular}{lccc}
\hline 
 & Normal & Intermediate & Superburst\tabularnewline
\hline 
Duration & $10$--$100\,\mathrm{s}$ & few minutes--hour & hours\tabularnewline
Fluence (erg) & $10^{39}$ & $10^{40}$--$10^{41}$ & $10^{42}$\tabularnewline
Recurrence time & hours--days & weeks--months & $1$~year\tabularnewline
Ignition column depth ($\mathrm{g\,cm^{-2}}$) & $10^{8}$ & $10^{9}$--$10^{10}$ & $10^{11}$--$10^{12}$\tabularnewline
Number observed & $>7000\,^{\mathrm{a}}$ & $\approx70\,^{\mathrm{b}}$ & $26\,^{\mathrm{c}}$\tabularnewline
Number of sources & $110$ & $\approx34$ & $15$\tabularnewline
Fuel & H/He & He & C\tabularnewline
\hline 
\end{tabular}
\par\end{centering}

$\,^{\mathrm{a}}$ MINBAR, \url{http://burst.sci.monash.edu/minbar}  \\
$\,^{\mathrm{b}}$ The long burst catalog assembled by the International Space Science Institute team on thermonuclar bursts (PI: A. Cumming), 
\url{http://www.issibern.ch/teams/ns_burster} \\
$\,^{\mathrm{c}}$ \cite[]{Zand2017}
\end{table}

\subsection{Intermediate duration bursts}

\label{sub:intermediate-duration-bursts}

\begin{figure}
\includegraphics[width=\textwidth]{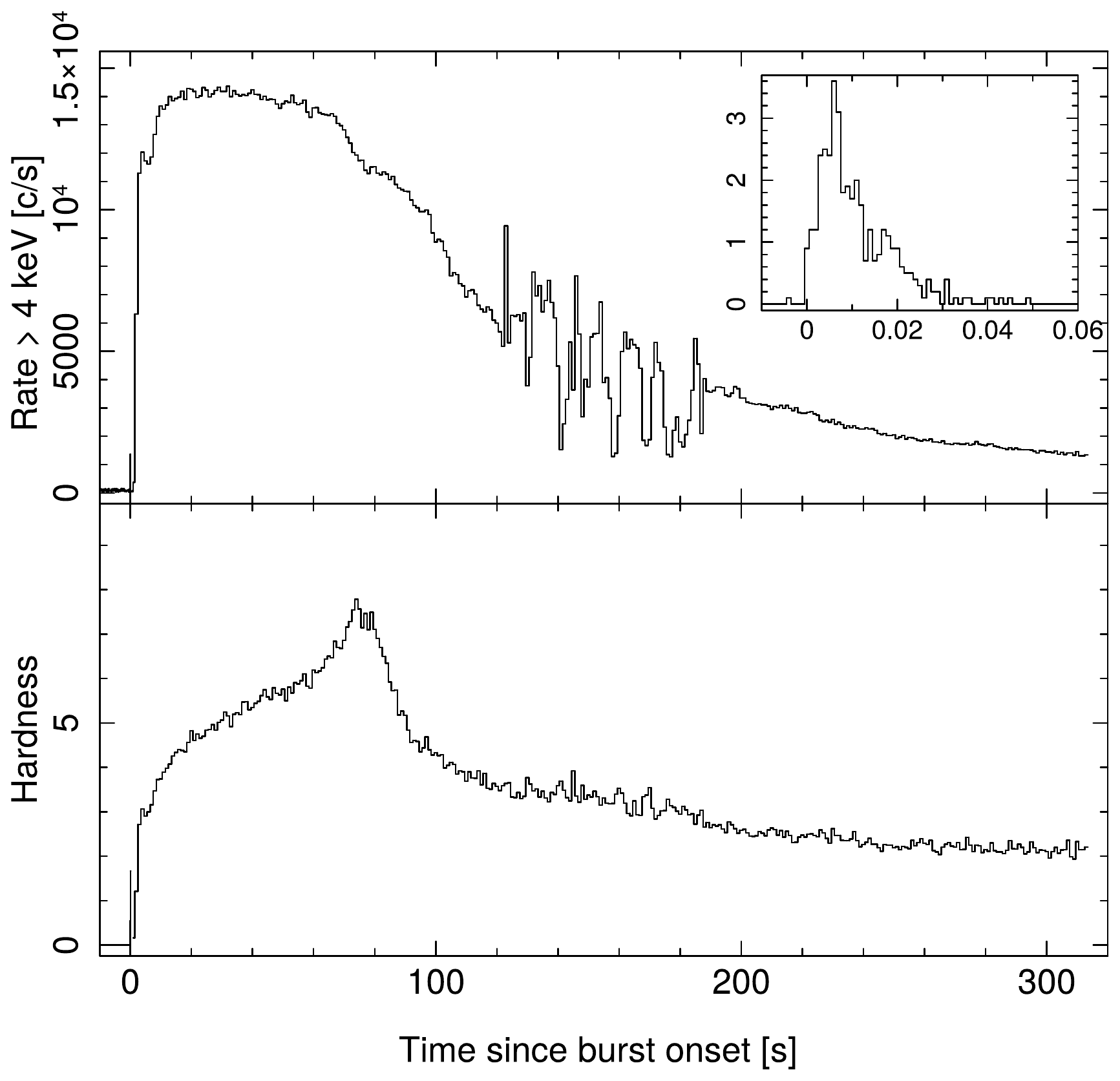}
\caption{Light curve of an intermediate duration burst from 2S\ 0918$-$549 observed with \xte/PCA. The observed count rate in the top panel exhibits strong
variability in the burst tail, which is achromatic in the PCA band pass, as it is not
apparent in the hardness (the ratio of the counts exceeding 4\ keV to those below; bottom
panel). The inset in the top panel zooms in on a short ``precursor'' at the
burst onset, marking the start of a brief period of superexpansion, during which the burst spectrum
temporarily drops below the PCA band. Although superexpansion lasts mere seconds, photospheric
radius expansion continues up to the peak in the hardness (``touch down''). 
Adapted with permission from \cite{zand11a} \copyright\ ESO.}
\label{fig:intermediate_duration_burst}

\end{figure}

Intermediate burst durations have been observed with exponential decay
timescales of several minutes up to $\sim 40$ minutes. These events are thought 
to be powered by deep helium ignition. 
At low mass accretion rates of $\approx0.001$--$0.01\,\dot{M}_{\mathrm{Edd}}$,
the neutron star envelope is relatively cool,
allowing a large
helium pile to accumulate prior to ignition. Furthermore, at low temperatures,
the ignition depth strongly increases with temperature (Fig.~\ref{fig:burning_regimes}),
such that small temperature variations produce bursts with a large
range of ignition depths and durations.  The durations can be anywhere
between a few minutes to tens of minutes, and may even rival the hours-long
superbursts (\S\ref{sub:Superbursts}). In a few cases of very low persistent
flux, the burst decay could be observed for several hours \cite{Degenaar2011,Keek2016igr1706}.
Observationally, 
the start of these longer events often falls in a data gap, so that the burst
duration and fluence are only known approximately, making it difficult
to accurately classify them. These events are observed both from sources where the
accretion composition is helium-rich \cite{Kuulkers2010} and 
hydrogen-rich \cite{Falanga2009}. In the latter case, hydrogen
burns at a shallow depth either stably (regime III in Table~\ref{tab:burning_regimes}) or in
weak flashes (regime II), producing a deep layer of helium.

As in all energetic helium-rich bursts, the burst flux  reaches the Eddington
limit, causing PRE (Fig.~\ref{fig:intermediate_duration_burst}). Often the radius expansion is particularly strong,
producing so-called ``superexpansion'' where the photospheric radius increases
by a factor $\sim10^{2}$. For two intermediate duration bursts, expansion
velocities of up to $30\%$ of the speed of light were inferred \cite{Zand2014}.
In the tail of some intermediate duration bursts strong flux variability
is observed (Fig.~\ref{fig:intermediate_duration_burst}), which may indicate interaction between the burst and
the accretion environment (see also \S\ref{sec:interaction}). Finally, convection at the onset of these powerful
bursts can potentially mix heavy ashes into the photosphere. Combined
with their relatively long duration, this makes intermediate duration
bursts the prime candidates for detecting redshifted spectral features
from the neutron star surface to constrain the dense matter EOS (\S\ref{sec:features}).

\subsection{Superbursts}

\label{sub:Superbursts}

X-ray bursts that last between several hours and a day are termed ``superbursts''.
Their duration corresponds to the cooling timescale at a column depth
of $10^{11}-10^{12}\,\mathrm{g\,cm^{-2}}$. At a mass accretion rate
of $0.1\,\dot{M}_{\mathrm{Edd}}$, it takes roughly a year to accumulate
such a layer. During that time all superbursting sources exhibit short
bursts. Therefore, the freshly accreted hydrogen and helium burns
at a smaller depth, and superbursts ignite in the carbon-rich ashes (Fig.~\ref{fig:burning_regimes}).
Most superbursts are observed from sources with accretion rates of
the order of $0.1\,\dot{M}_{\mathrm{Edd}}$, but exceptional cases
are known: 4U~0614+09 with $\sim0.01\,\dot{M}_{\mathrm{Edd}}$ exhibited
two superbursts \cite{Kuulkers2010,Serino2016}, and GX~17+2
at $\sim1\,\dot{M}_{\mathrm{Edd}}$ showed four \cite{Zand2004}.
Because their duration is often much longer than the typical satellite (low-Earth) orbit of $\approx90\,\mathrm{min}$, the start is in many cases obscured by Earth occultations.
Combined with the sparse sampling and limited data quality of all-sky
instruments, this limits in many cases our ability to measure the
duration and fluence of the event, and we refer instead to \emph{candidate}
superbursts. The long recurrence time makes superbursts the rarest
burst category: since their discovery 
\cite{corn00,stroh02}
only 25 (candidates) have been observed (see \cite{Zand2017} for a
recent review and list of detections).

In the rare cases when the start of a superburst was observed, a short
precursor is visible directly prior to the superburst. After the carbon
ignition $\sim100\,\mathrm{m}$ below the neutron star surface, a
sound wave may travel toward the surface ahead of the carbon flame
\cite{Weinberg2006sb,Weinberg2007,keek11,keek12a}. 
As it propagates
to lower density regions, it may speed up and exceed the speed of
sound, turning into a shock. When it reaches the surface, the kinetic
energy of the shock heats the outer layers, producing a short X-ray burst.
Furthermore, any hydrogen or helium present there will be burned,
adding to the energy of the precursor burst. Only in one observation
could the precursor's fluence be accurately determined: it was 40--100\% larger than the fluence of the regular short helium bursts observed
from this source, indicating that another source of energy contributed,
such as shock heating \cite{Keek2012precursors}.

Following the precursor, the rise of a superburst light curve typically
is rather slow: in 2001, the superburst flux from 4U~1636$-$536 reached
its peak only after $\sim900\,\mathrm{s}$ (Fig.~\ref{fig:examples} bottom). With the exception of the
1999 superburst from 4U~1820$-$20, which reached the Eddington limit
\cite[]{stroh02}, most
superbursts likely have such a slow rise. The shape of the light curve is attributed
to the cooling of an envelope with a particular temperature profile
\cite{keek15a}. This profile is left behind by the radially outward
moving carbon flame. Fits with cooling models measure a profile close
to $T\propto y^{1/4}$. This suggests that as the carbon flame moved
into lower density regions near the surface, the fraction of carbon
that burned decreased, and the flame likely stalled before reaching
the photosphere.

After the superburst has decayed, it continues to have an effect on the neutron star envelope
by quenching the occurrence of regular bursts for several weeks \cite{Kuulkers2003a}. Heat deposited by
the superburst in the envelope causes freshly accreted hydrogen and helium to burn
in a stable manner (regime VII in Table~\ref{tab:burning_regimes}). Even after the
superburst has decayed, the stable burning can sustain itself for several weeks.
Once the envelope has cooled down sufficiently, the burning mode transitions back
to producing bursts \cite{keek12a}. 

Superbursts ignite close to the outer crust,
and are sensitive to the 
thermal properties
at that depth. This region is rather poorly understood, with observations
of cooling quiescent sources suggest the presence of an unknown heat
source \cite{bc09,Deibel2015}, and where Urca neutrino cooling may play
an important role in regulating the temperature \cite{Schatz2014Nature}.
In recent years superburst candidates have been detected from transient
sources \cite{Keek2008,Serino2016}. These observations pose several challenges. First, the
persistent flux may evolve on similar timescales to a superburst,
and it is difficult to make the definite determination of the thermonuclear
nature of the event. Second, the superbursts occur within months or
sometimes days of the start of an accretion outburst \cite{Altamirano2012,Serino2012}. Current theory
predicts that the outer crust of the neutron star is not heated sufficiently
at that time to ignite a carbon flash. Yet, superbursts are observed,
suggesting the presence of an unknown heat source near the superburst
ignition depth. This may be related to the shallow heat source inferred
for cooling quiescent sources \cite{Deibel2016}.

It is challenging to explain the origin of the carbon fuel. Comparison
of the X-ray light curves to cooling models finds that the carbon
mass fraction of the superburst fuel is $15-30\%$ \cite{cumming06}.
The rest of the fuel may be iron and heavier elements produced during
the {\sl rp}-process. If a substantial fraction is heavy isotopes
near the end point of the {\sl rp}-process, their photodisintegration could
account for as much as $50\%$ of the superburst energetics \cite{Schatz2003ApJ}.
However, simulations typically predict most heavy isotopes to be near
the iron-group \cite{woos04,Jose2010}.
Carbon is typically not accreted in substantial quantities,
and therefore has to be produced by nuclear burning of the accreted
hydrogen and helium. $3\alpha$ burning of helium during ``normal'' (mixed H/He) bursts produces
carbon, but the high temperatures reached during these events enable $\alpha$- and
proton-captures also to destroy carbon. Detailed simulations predict that
the net carbon production by normal bursting activity is at most a mass fraction of
$\approx5\%$ \cite{woos04}. Stable burning of hydrogen and helium
can produce large quantities of carbon \cite{Stevens2014}, but this
burning has also been thought to only take place at high accretion rates, where the temperature
in the envelope again is high enough to allow for the reactions that
destroy carbon. A new stable regime has recently been identified in
simulations \cite{Keek2016stable}, where stable burning produces
copious quantities of carbon at lower mass accretion rates near $0.1\,\dot{M}_{\mathrm{Edd}}$,
similar to where most superbursts are observed. All superbursting
sources also exhibit short hydrogen or helium flashes, but they have
a high value of the $\alpha$-parameter \cite{Zand2003}. This suggests
that a substantial part of hydrogen and helium are burned in a stable
manner in between the short bursts. Future studies will need to determine
if this involves the newly suggested stable regime as a source of
the carbon fuel for the superbursts. 

In the absence of spectroscopic information, superbursts provide
the only observational constraints on the composition of the ashes
of hydrogen/helium burning in neutron star envelopes. Furthermore,
superburst ashes are compressed and form the crust. These ashes consist
predominantly of $^{56}\mathrm{Fe}$ created in nuclear statistical
equilibrium, where any heavy {\sl rp}-process isotopes
have been destroyed by photodisintegration.

\begin{figure}
\includegraphics[width=0.46\textwidth]{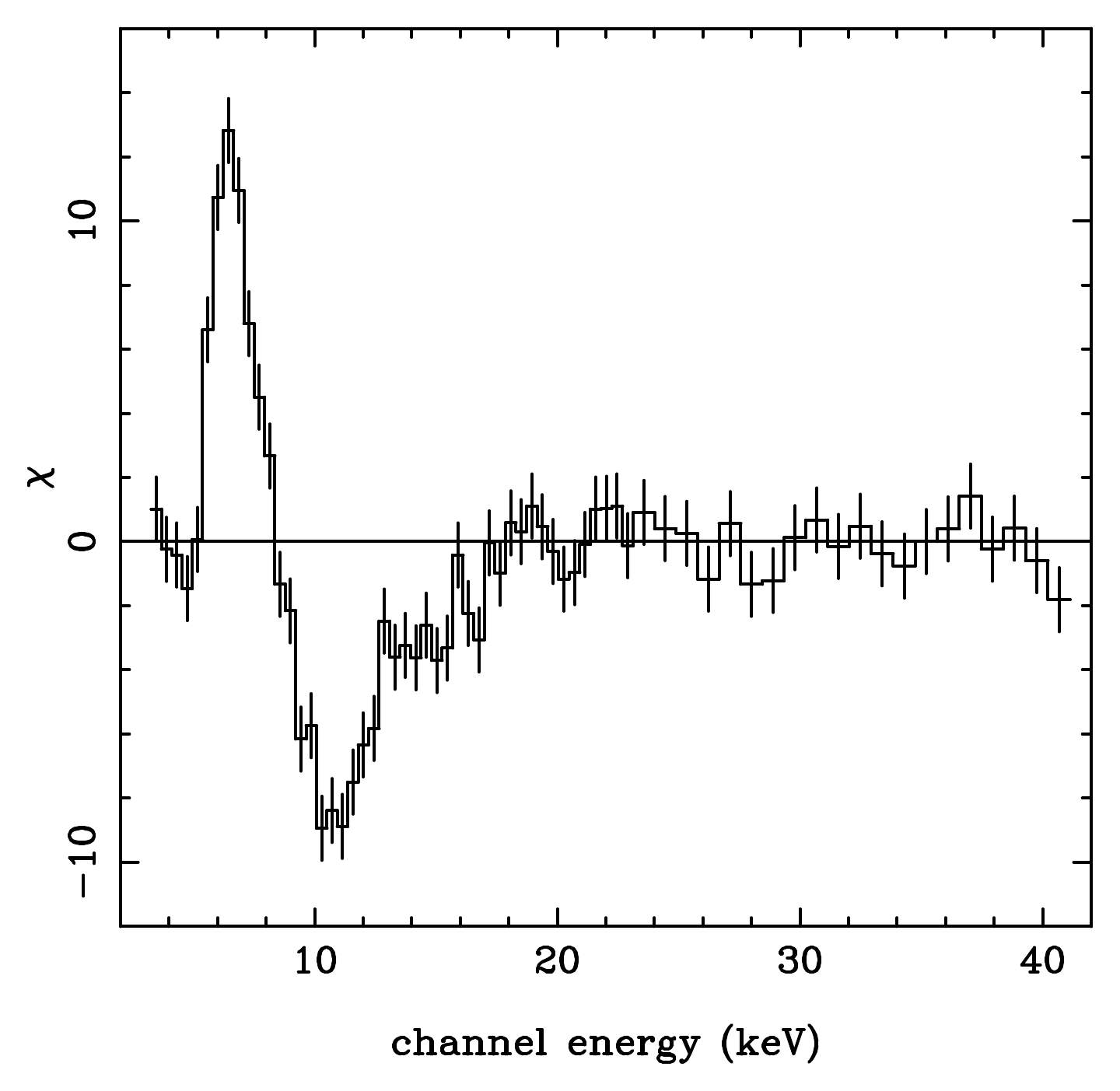}
\includegraphics[width=0.54\textwidth]{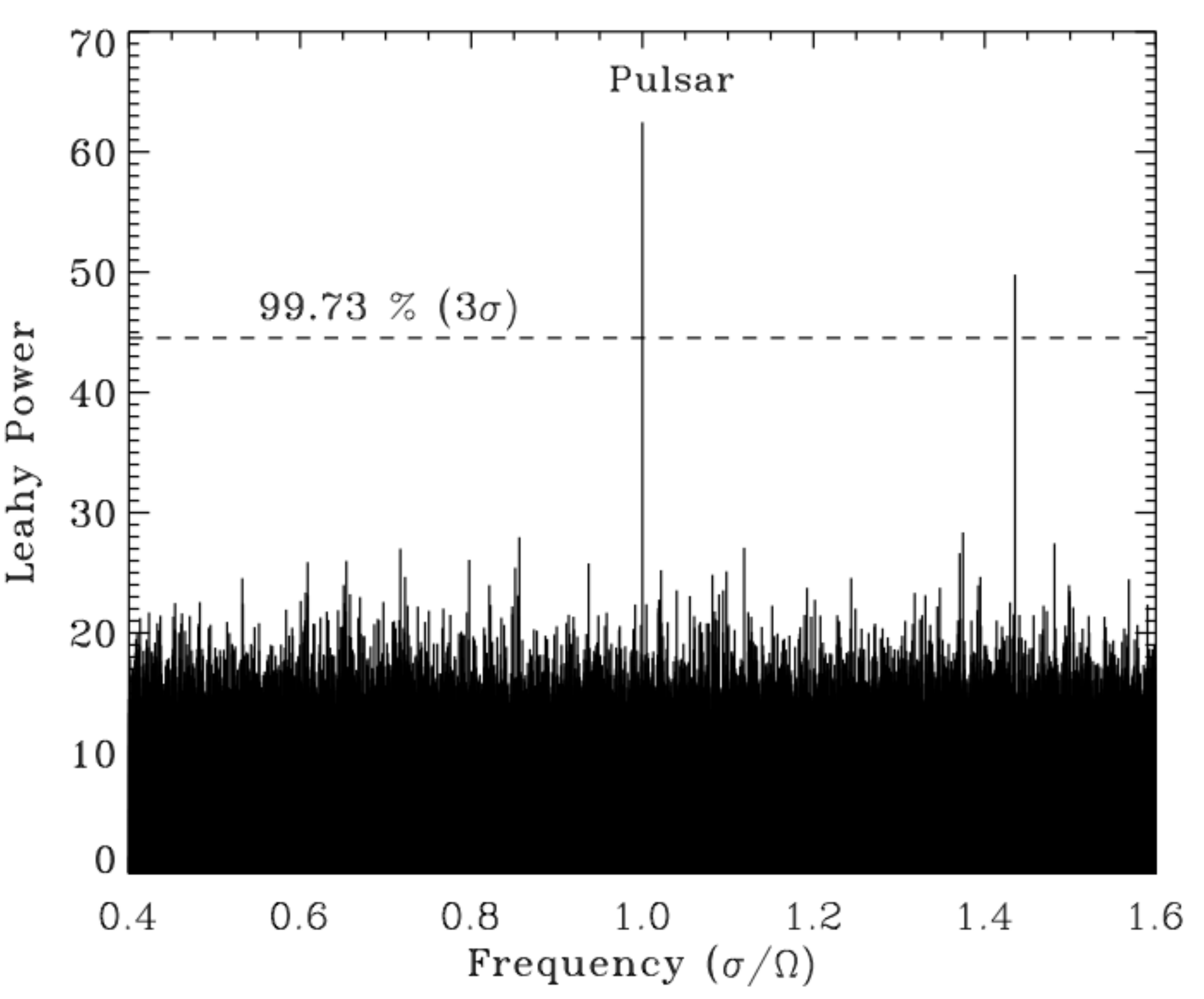}
\caption{Detailed spectroscopy and timing studies of the superburst from
4U~1636$-$536 observed in 2001 with \xte/PCA (see also Fig.~\ref{fig:examples} bottom).
Left: after subtraction of the thermal continuum spectrum, the iron emission line and absorption edge
are visible as signatures of reflection off the disk (from \cite{Strohmayer2002}).
Right: power spectrum during the burst showing pulsations  detected at the neutron star spin frequency, $\Omega$, as well as another
mode of unknown origin at higher frequency (from \cite{Strohmayer2014}).}
\label{fig:superburst_features}

\end{figure}

The long durations of superbursts allow for detailed spectra to be
obtained compared to short bursts. Most detections are, however, with
instruments that have a relatively small effective area and a high
background. The highest quality spectra were obtained for two superbursts
with the RXTE/PCA: 4U~1820$-$30 in 1999 and 4U~1636$-$536 in 2001. The
spectra exhibited photoionized reflection of the burst off the accretion
disk (Fig.~\ref{fig:superburst_features}, left) and a variable persistent
component (see also \S\ref{sec:interaction}).
A detailed timing study also  revealed
pulsations at the neutron star spin frequency, modulated by Doppler shifts
from the binary orbit, as well as a (presumed) global oscillation mode at higher frequency
(Fig.~\ref{fig:superburst_features} right).

\section{Thermonuclear burst simulations}
\label{sec:simulations}

Simulating X-ray bursts is a problem with extreme scales. The strong
dependence of the nuclear reaction rates on temperature demands time
steps of less than a nanosecond, whereas the recurrence time of bursts
is hours and that of superbursts is typically a year or longer. Resolving
convective mixing at the onset of a burst requires a spatial resolution
of half a centimeter \cite{Malone2011}, whereas a flame spreading
around the star travels over $30\,\mathrm{km}$. Furthermore, for
each zone of the model thousands of nuclear reactions have to be evaluated
among hundreds of isotopic species. Numerical simulations of X-ray
bursts, therefore, need a large number of time steps and zones as
well as an extensive nuclear network, a computational task which is
not yet feasible on present-day hardware. All current simulations
reduce the problem. 2D and 3D models explicitly model turbulent mixing,
but typically simulate only part of a burst with a small selection
of nuclear reactions to limit the computational expense. On the other
hand, one-dimensional simulations include an approximation of turbulence,
but this allows them to use an implicit scheme and calculate series
of multiple complete bursts. Moreover, the number of zones in such a model
is small enough to use a full nuclear network. Here we discuss the
progress in recent years for the different categories of models.

\subsection{Single-Zone Models}

The largest simplification of any simulation is the one-zone model,
where the entire neutron star envelope is reduced to a single zone.
Time evolution can be implemented by considering the changes in the
the fuel column due to accretion and nuclear burning. Such simplification
is helpful when illustrating the different regimes of nuclear burning
(\cite{Fujimoto1981,Heger2005}; see also Fig.~\ref{fig:burning_regimes}). Single-zone
models, however, miss several important effects from nuclear burning
taking place simultaneously at different depths or from the mixing
of the fuel and ashes layers. That behaviour can only be resolved with
multi-zone models.

By considering only a single zone, most computational resources can
be applied to large nuclear networks. Such models are used to evaluate
the impact on X-ray burst light curves and burning ashes of uncertainties
in a large number of reaction rates \cite{Iliadis1999,Parikh2008,cyburt16}.
Whereas some one-zone models include the thermodynamic response of the neutron atmosphere to the nuclear burning, others rely on the ignition conditions
or even full thermodynamic trajectories from multi-zone models. The latter
is not a good approach for studying X-ray bursts, because changes
to any rate potentially produce a significant change in the temperature,
invalidating the trajectory. Such changes to the trajectory are difficult
to predict a priori, as the nuclear flow may progress through different
branches, and a change in one reaction may cause the nuclear flow
to be rerouted through another path. The collective effect of the
reactions along the altered path can significantly affect the ashes
composition and the observed light curve. Therefore, an important
aspect of evaluating a nuclear reaction rate is its impact on \emph{changing}
the thermodynamic trajectory. Given that multi-zone models can now
be run relatively fast on current hardware, the use of trajectories
should be avoided, and the only remaining role for one-zone models
may be in exploratory studies of large parameter spaces, to be followed
up with multi-zone models \cite{cyburt16}.

\subsection{One-Dimensional Multi-Zone Models}

\begin{figure}
\includegraphics[width=\textwidth]{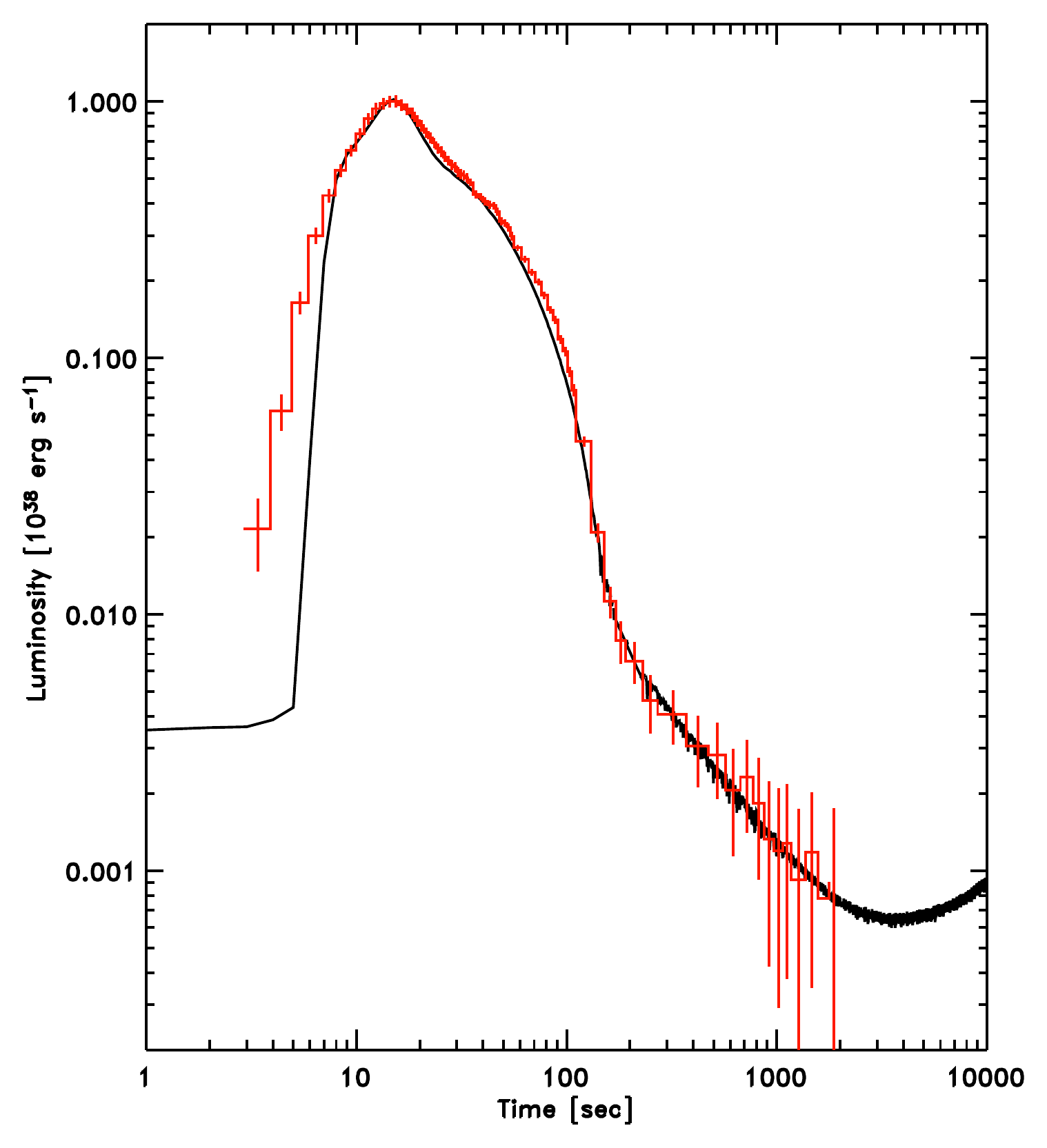}
\caption{Comparison of a burst observed from GS~1826$-$24 with \xte/PCA
(red histogram with error bars) and a one-dimensional multi-zone simulation
with KEPLER (black line). The model of a mixed
hydrogen/helium burst (regime V in Table~\ref{tab:burning_regimes}) reproduces
the burst shape well. The light curve from the peak to $\approx 120$~s is shaped
by the {\sl rp}-process, which is modeled in full detail. Afterwards, the
flux decay follows the cooling of the neutron star envelope, which can be
traced in the observation for over $1000$~s. The observed rise is slower than
predicted: this may be the effect of flame spreading, requiring multi-dimensional
models.
Adapted with permission from \cite{Zand2009} \copyright\ ESO.}
\label{fig:kepler1826}

\end{figure}

One-dimensional models resolve the neutron star envelope into multiple
zones in the radial direction, usually on a Lagrangian grid which
employs mass coordinates. Most simulations use an implicit hydrodynamics
scheme, which assumes hydrostatic equilibrium and uses mixing length
theory to approximate convection and other turbulent mixing processes.
This allows for the model to be evolved over long times, simulating
multiple complete bursts. The past decade saw the introduction of
full nuclear reaction networks covering up to $\approx1300$ isotopes
\cite{woos04,Fisker2006,Jose2010} (Fig.~\ref{fig:nuclear_network}). These models are being used
to study a wide range of bursts and burning phenomena \cite{Lampe2016}.
A major success of these models is their good agreement with the observed
light curve of bursts from the ``Clocked Burster'', GS~1826-24 \cite{Heger2007,Zand2009} (Fig.~\ref{fig:kepler1826}). 

The implementation of large nuclear networks confirmed that approximated
networks were responsible for irregular bursts in past models, especially
with respect to approximations of the {\sl rp}-process \cite{taam93}:
full reaction networks are required to 
simulate behaviour consistent with observations \cite{woos04}.
Moreover, interactions between different zones influence the nuclear
burning. Ashes of the previous burst are mixed with fresh fuel, and
typically shorten the waiting time for the next burst to ignite. This effect
is referred to as ``compositional inertia'' \cite{taam93,woos04,Jose2010}.
Bursts with a smaller ignition column depth are less powerful, reach
lower peak temperatures, and the nuclear flow 
does not extend as far along
the {\sl rp}-process path. Therefore, compositional inertia changes the
composition of the  ashes that fuel superbursts and forms the outer crust. 
Although it is typically not taken into account self-consistently
in burst simulations, nuclear heating and cooling processes in the
crust depend very sensitively on the ash composition \cite{Schatz2014Nature,Deibel2016}.
Furthermore, as many of the required nuclear reaction rates are highly
uncertain, extensive studies have been undertaken to find the most influential
reactions and to quantify their effect on burst observables \cite{Iliadis1999,Parikh2008,cyburt16}.
The CNO breakout reaction $^{15}\mathrm{O}(\alpha,\gamma)^{19}\mathrm{Ne}$
is found to be particularly important for the stability of the burning
processes (\cite{Fisker2006,Fisker2007,Davids2011,keek14b,cyburt16}; see also \S\ref{sec:regimes} and \S\ref{subsec:msb}).

The burning processes powering pure helium bursts were thought to
consist of $3\alpha$ and a straightforward series of $\alpha$-captures.
Simulations with full networks find, however, that $(\alpha,p)$ reactions
can produce a small number of protons. The protons act as a catalyst
and substantially speed up the burning process by by-passing $^{12}\mathrm{C(\alpha,\gamma)^{16}O}$
through the faster reactions $^{12}\mathrm{C(p,\gamma)^{13}N(\alpha,p)^{16}O}$.

In recent years multi-zone models of deep carbon burning have been
created to model superbursts \cite{Weinberg2006sb,Weinberg2007,keek12a}.
These models directly accrete carbon-rich material to avoid the computational
expense of simulating $\sim10^{3}$ hydrogen/helium flashes, as well
as to avoid the problem that models of hydrogen/helium typically do
not produce sufficient carbon. In some studies a hydrogen/helium atmosphere
is added to the model close to the moment of ignition, such that the
effect of the deep carbon flash on the outer atmosphere can be investigated.
Hydrogen/helium burning at the superburst onset contributes to the
precursor burst \cite{Weinberg2007,keek12a}. Furthermore, the hot
ashes of the superburst cause freshly accreted hydrogen and helium
to burn stably, quenching short X-ray bursts for days to weeks \cite{Kuulkers2003a,keek12a}.
An unsolved issue is that the carbon burning propagates radially as
a convective flame, but convection is not properly modeled by implicit
codes. The one-dimensional simulations find the flame to spread supersonically
as a detonation \cite{Weinberg2006sb}, whereas multi-dimensional
models of comparable carbon flames in Type Ia supernovae produce deflagrations
\cite{Woosley2011}.

Multi-zone models have been very successful at simulating a wide range
of observed bursts and other burning behaviour (e.g. \cite[]{lampe16}). It has proven a challenge,
however, to reproduce the observed conditions where the different
burning regimes occur. In part this is due to the large parameter
space that needs to be investigated because of uncertainties in, for
example, the accretion composition and key nuclear reaction rates.
Furthermore, heating and cooling processes in the crust have a strong
effect on the burning behaviour \cite{keek09,keek11,Zamfir2014},
but their influence has been explored in a limited way thus far. The 
same holds true for rotationally induces turbulent mixing \cite{Piro2007,keek09}.

\subsection{Multi-Dimensional Models}
\label{sec:multid_models}

\begin{figure}
\includegraphics[width=\textwidth]{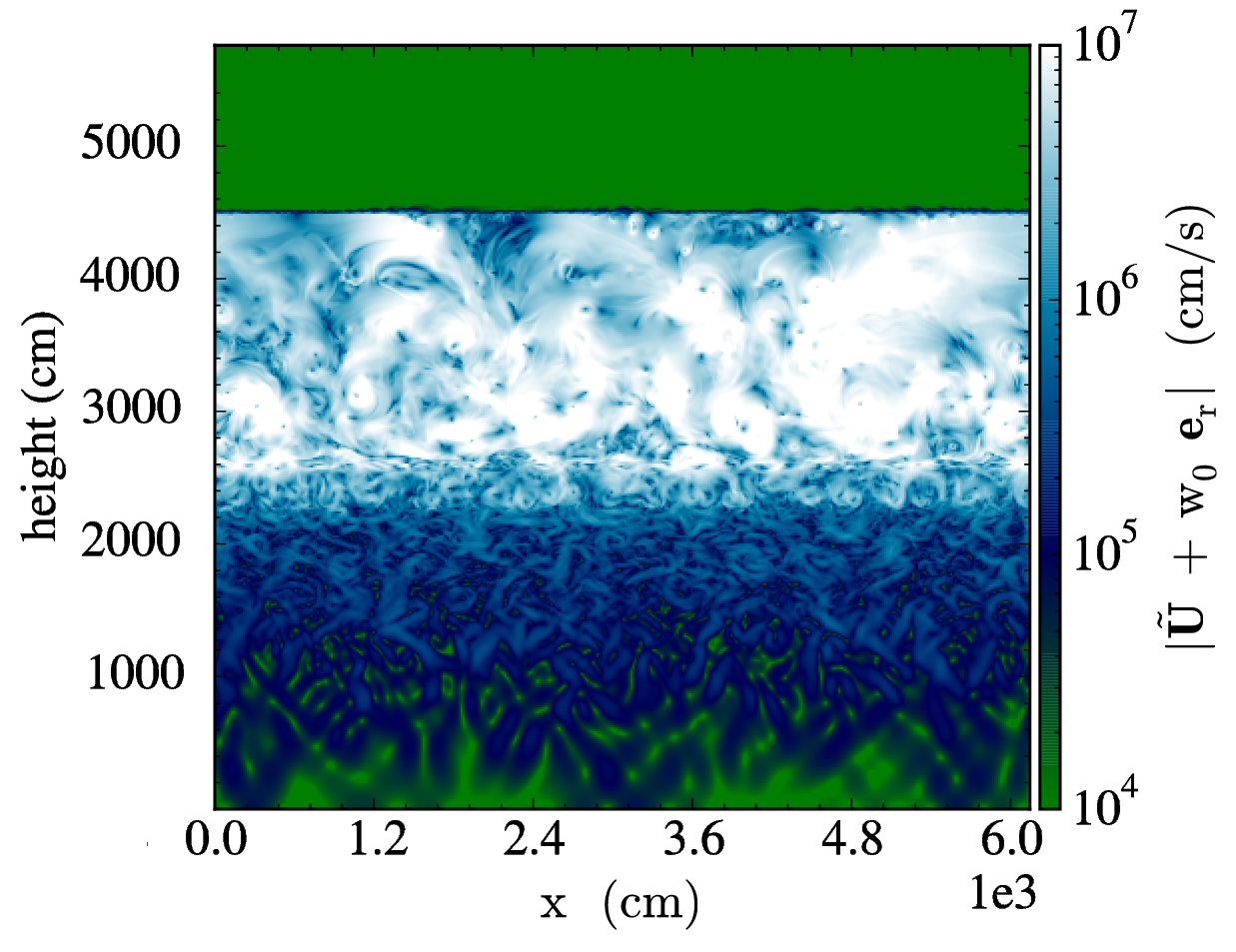}
\caption{Magnitude of the velocity field in a two-dimensional hydrodynamics
simulation with MAESTRO. Due to the large amount of
heat generated by the thermonuclear runaway, convection is initiated. It
dominates the energy transport shortly after the burst onset, and mixing of
the composition may deposit burst ashes close to the surface. To correctly model
convective mixing during an X-ray burst, a spatial resolution of $0.5$~cm is required 
\cite{Malone2011}. Adapted from \cite{Zingale2015}.}
\label{fig:multid}

\end{figure}

Multi-dimensional models of X-ray bursts either simulate a small box
in the neutron star envelope aiming to resolve convection, or they
simulate flame spreading accross the entire envelope with a cruder
implementation of turbulent mixing. The earliest efforts in the former
category are simulations of deep helium ignition \cite{Zingale2001,Simonenko2012},
which may be applicable to the most powerful intermediate duration
bursts. For pure helium bursts, a small nuclear network may suffice.
Furthermore, these simulations find that at large depths, the fast
helium flame travels as a detonation, such that only a short simulation
time is required. Fast flame propagation could be a requirement for
explaining exceptionally fast rise times observed in the most powerful
helium bursts \cite{Zand2014}. Recognizing that most observed bursts
ignite at a shallower depth and thus propagate more slowly as a deflagration,
a new ``low Mach number'' code has been developed to make simulations
more efficient. The first simulations with the MAESTRO code have explorered
the resolution requirements, and find that a zone size of $0.5\,\mathrm{cm}$
is required to resolve convective mixing during the burst \cite{Malone2011,Malone2014,Zingale2015}
(Fig.~\ref{fig:multid}).
Potentially, such simulations can be used to calibrate the approximation
for mixing in the one-dimensional codes.

At the onset of a burst, the flame travels a longer path than what
can be fully resolved by current simulations. Two-dimensional simulations that
include the radial and latitudinal (or longitudinal) directions, adopting zones that are
elongated in the latter dimension. The zones are not fine enough to resolve
the small-scale turbulence during the convective stage at the burst
onset, but they are sufficient for modeling the larger scale mechanisms
that drive the flame propagation. These simulations find that initially,
the flame must be confined by the Coriolis force and/or magnetic tension in order to prevent
it from 
dying out
\cite{slu02,cavecchi13,cavecchi15,Cavecchi2016}.
Subsequently, 
a combination of the Coriolis force, magnetic tension-induced friction in the sheared burning front, and conduction causes
the flame to spread accross the surface. 

\section{Nuclear experimental physics}
\label{sec:nuclear}

\begin{figure}
\includegraphics[width=\textwidth]{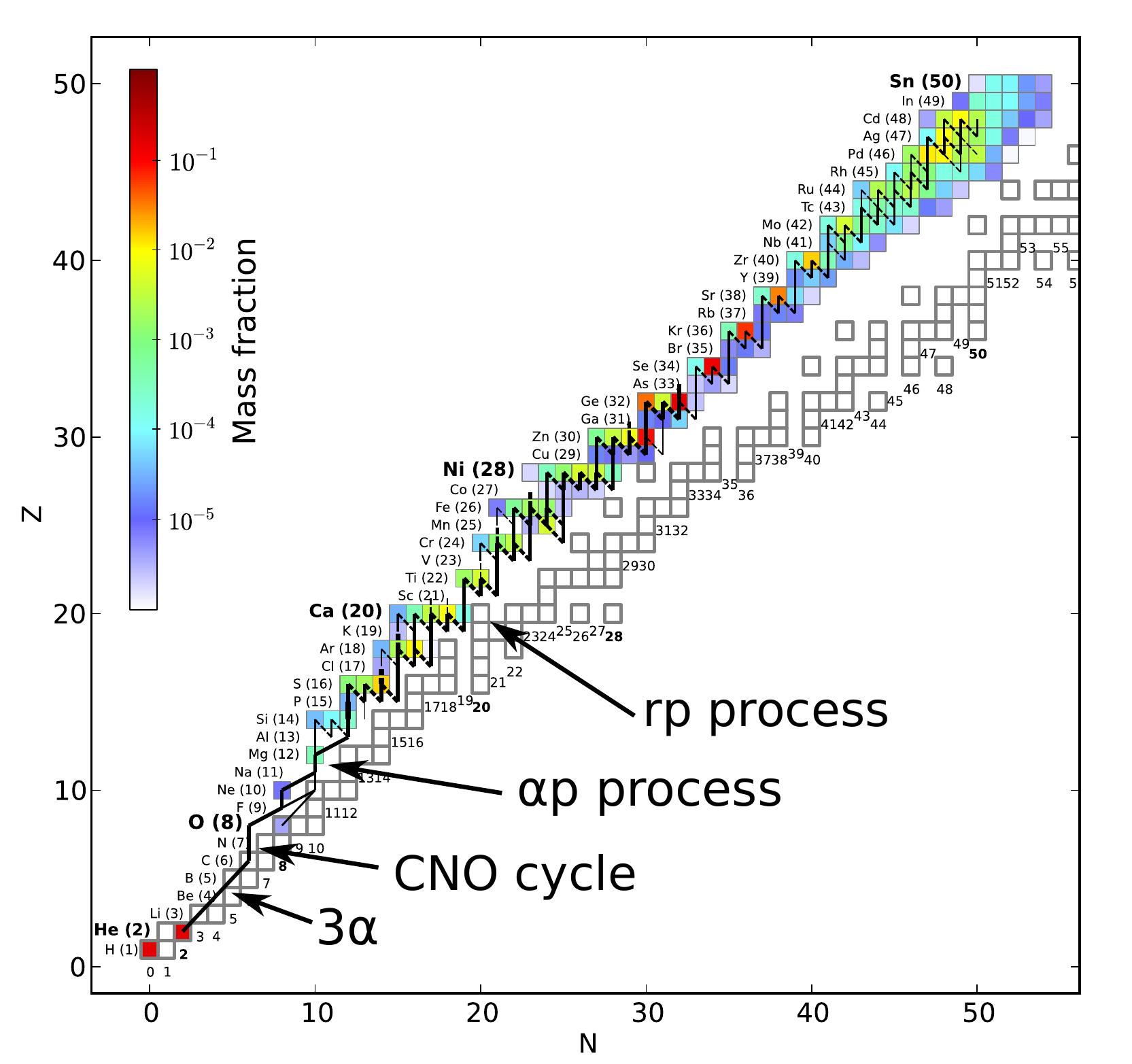}
\caption{Flow of nuclear reactions between isotopes with neutron number N and proton
number $Z$, where thicker lines represent larger net flows; from a KEPLER simulation 
with a solar accretion composition \cite{keek12a} approximately a millisecond into the
burst. Colours indicate the mass fraction of the isotopes, and a thick border marks
the stable isotopes. Helium burns via the $3\alpha$ process, followed by hydrogen burning in the 
$\beta$CNO cycle. Break-out from the cycle flows into the {\sl$\alpha$p} process and
subsequently the {\sl rp}-process, which extends up to tin (Sn). At this stage of
the burst, most isotopes are unstable, meaning that after the burst they $\beta$-decay
to form stable isotopes. The most abundant isotopes along the {\sl rp}-process path
are at ``waiting points'', 
such as $^{64}$Ge, $^{68}$Se and $^{72}$Kr, where reaction products accumulate due to slow subsequent $\beta$-decays.
}
\label{fig:nuclear_network}

\end{figure}

A significant fraction of the research effort directed at thermonuclear bursts over the last few decades has been in the area of nuclear experimental physics. The nuclear reactions that may take place in bursts, where the temperature can reach $10^8-10^9$~K, were first explored by \cite{ww81}. The most important reactions are the $\beta$-limited CNO cycle, in systems that accrete hydrogen, and the triple-$\alpha$ reaction, which fuses helium to carbon. 

The triple-$\alpha$ reaction first produces carbon, which (via the CNO cycle) catalyses the burning of hydrogen. CNO breakout reactions including $^{15}\mathrm{O}(\alpha,\gamma)^{19}\mathrm{Ne}$ then lead to a series of $(\alpha,p)$ and $(p,\gamma)$ reactions (the {\sl $\alpha$p}-process), feeding into the 
rapid-proton, or {\sl rp}-process. The {\sl rp}-process was identified as a key reaction chain that could produce  elements far heavier than any of the precursor reactions. 
However, experimental data for the neutron-deficient nuclei involved was lacking for many years, forcing modelers to rely on theoretical calculations for masses and rates \cite{schatz98}.
At the same time, it was initially unclear how far the {\sl rp}-process reactions extended. Early simulations were based on limited reaction networks, under the assumption that further proton captures beyond $\nucl{56}{}{Ni}$ could be neglected. Such simulations also tended to find non-negligible reaction products accumulating at the limits of the reaction rate networks. A simulation study using a one-zone model determined the ultimate limit of {\sl rp}-process burning in the Sn-Sb-Te cycle  \cite{schatz01} (see also \cite{Koike2004}). This result implies that {\sl rp}-process reaction products are limited to atomic number $Z \leq 54$, and that  the ashes forming the crust are made up of nuclei lighter than mass number $A \approx 107$.

With the maximum extent of the {\sl rp}-process burning now constrained, the goals of experimenters became to improve the mass measurements for proton-rich nuclei (e.g. \cite{so17}), as well as determining the key reactions that might influence the properties of the bursts. 
The experimental studies informed, and were further motivated by, development of more accurate numerical models of bursts, as described in \S\ref{sec:simulations}. 
Simulations enabled a very detailed picture of the nuclear burning processes during a burst to emerge (e.g. \cite{fis08}), although the fidelity of this picture remains uncertain.

Any individual reaction will have little impact on the burst profile unless a significant fraction of the fuel mass is processed through that reaction, and it's timescale is unusually slow \cite{vanwormer94}.
Substantial experimental work has established several such
``waiting points'', including $\nucl{60}{}{Zn}$, $\nucl{64}{}{Ge}$, $\nucl{68}{}{Se}$ and $\nucl{72}{}{Kr}$ 
(e.g. \cite[]{winger93a,blank95,pfaff96}).
The effect of the reactions which produce these nuclei is illustrated in Fig.~\ref{fig:nuclear_network}, with a substantial fraction of the mass flow ``piling up'' in these isotopes.
Further proposed waiting points, of $\nucl{22}{}{Mg}$, $\nucl{26}{}{Si}$, $\nucl{30}{}{S}$, and $\nucl{34}{}{Ar}$ were suggested as the explanation for seldom-observed bolometrically double-peaked bursts \cite{fisker04}. 
However, an explanation for these events in terms of nuclear physics must also necessarily explain why the double-peaked behaviour is not observed in every burst.

More systematic investigations using a combination of single- and multi-zone models attempted to identify all of the most significant reactions that might be expected to affect the burst lightcurves \cite{cyburt10}. These investigations were integrated into  the ongoing {\sc reaclib}\footnote{{\url https://groups.nscl.msu.edu/jina/reaclib/db}} project, which seeks to collect, curate, and distribute data on nuclear reaction rates and masses to broadly impact the astrophyisical modelling community. The most recent results from this project have provided a clear list of priority reactions for investigation via nuclear experiment \cite{cyburt16}. However, it is worth noting that the sensitivity studies are based on a single burst model, and it is possible that different reactions may have significant impacts on the lightcurves of bursts in different regimes (see \S\ref{sec:regimes}).

A detailed description of the nuclear experiments focussed on reactions or masses of relevance for thermonuclear bursts is beyond the scope of this review (see instead \cite{Parikh2013}). However, we comment that in the past few years an increasing level of interaction has taken place between the experimental community, and burst observers and modellers, enabled in a large part by the activities of the Joint Instutute for Nuclear Astrophysics: Centre for the Evolution of the Elements (JINA-CEE\footnote{\url{http://jinaweb.org}}). This growing dialogue, in combination with new experimental techniques, and upcoming hardware including the Facility for Rare Isotope Beams (from 2021 onwards), suggest that the most exciting time for studying the nuclear reactions in bursts is yet to come.

\section{Summary and outlook}

In the last decade, studies of thermonuclear bursts have entered an exciting new phase, where the interplay of observations, numerical modelling, and nuclear experiments are  making substantial progress on understanding the burst phenomenology. At the same time, the interpretation of observational data (motivated in part by constraining the mass and radius; see 
\cite[]{miller13}) has driven new investigations into the spectral formation processes.

The diversity of  burst behaviour, although still not completely understood, presents a continuing challenge to these investigations.
The computational challenges in simulating the entire neutron star surface remain, but even 1-D models are achieving success in reproducing an increasing range of observed phenomena.
Assembly and analysis of large databases of observations and simulations appear to offer the best chance for  understanding all the possible ignition cases.
More comprehensive sensitivity studies have the prospects of conclusively identifying the nuclear reactions most important to the full range of burst types, and hence directly motivating future experimental efforts.
Thermonuclear burst physics has been identified as a substantial priority for future nuclear physics experiments \cite{nucastro16}.

New observations are continually being sought, with the available fleet of long-duration X-ray missions in addition to a handful of recently-launched instruments. 
The current prospects for detection of new examples of burst oscillations appear excellent, with the 2015 launch of India's {\it ASTROSAT} mission, featuring a large-area proportional counter with similar capabilities to \xte's PCA. Early observations of 4U~1728$-$34 have detected high-frequency varability as seen earlier with \xte{} \cite{vc17}.
The 2017 deployment of the {\it NICER} instrument to the International Space Station offered another avenue for detection of high-frequency variability, and the ``first-light'' observations have detected burst oscillations in 4U~1608$-$52\footnote{\url{http://www.nasa.gov/press-release/goddard/2017/nasa-neutron-star-mission-begins-science-operations}}.

These new data also offer the prospect of new insights, particularly where the observational capabilities exceed those to date.
In particular, there are exciting prospects of more detailed burst timing and spectral information from recently-launched large-area instruments including {\it NICER} and LAXPC, aboard {\it ASTROSAT}; and in the more distant future, 
{\it eXTP} and {\it Strobe-X}. The data from these instruments may more clearly distinguish the reflection spectrum, and perhaps also finally realise the promise of burst oscillations for precisely constraining neutron star mass and radius.

\begin{acknowledgement}
The authors are grateful for helpful comments from Y.~Cavecchi, M.C.~Miller, and
H.~Schatz.
This work was supported in part by the National Science Foundation under Grant No. PHY-1430152 (JINA Center for the Evolution of the Elements).
The authors are grateful for support received as part of the International Team on Nuclear Reactions in Superdense Matter by the International Space Science Institute in Bern, Switzerland.
\end{acknowledgement}

\newcommand*{\doi}[1]{\href{https://dx.doi.org/#1}{DOI~#1}} 
\urlstyle{same} 
\bibliographystyle{spphys}
\bibliography{all,lk} 

\end{document}